# PENCO: A Physics–Energy–Numerics–Consistent Operator for 3D Phase Field Modeling

Mostafa Bamdad[*1] | Mohammad Sadegh Eshaghi[†2] | Cosmin Anitescu[2] | Navid Valizadeh[2] |Timon Rabczuk[‡1]

[1] Chair of Computational Mechanics, Institute of Structural Mechanics, Bauhaus-Universität Weimar, 99423 Weimar, Germany

[2] Chair of Computational Science and Simulation Technology, Institute of Photonics, Department of Mathematics and Physics, Leibniz University Hannover, 30167 Hannover, Germany

**Abstract**

Accurate and efficient solutions of spatiotemporal partial differential equations (PDEs), such as phase-field models, are fundamental for understanding interfacial dynamics and microstructural evolution in materials science and fluid mechanics. Neural operators (NOs) have recently emerged as powerful data-driven alternatives to traditional solvers; however, existing architectures often accumulate temporal errors, struggle to generalize over long temporal horizons, and require large training datasets. To overcome these limitations, we propose PENCO (Physics–Energy–Numerics–Consistent Operator), a hybrid operator-learning framework that integrates physical laws with data-driven neural operator methods, using either the Fourier Neural Operator (FNO-4D) or the Multi-Head Neural Operator (MHNO) architecture as the backbone. The formulation introduces an enhanced $L^2$ Gauss–Lobatto collocation residual around the temporal midpoint that robustly enforces the governing dynamics and significantly improves accuracy, a Fourier-space numerical consistency term that captures the balanced behavior of semi-implicit discretizations, and an energy-dissipation constraint that ensures thermodynamic consistency. Additional low-frequency spectral anchoring and teacher-consistency mechanisms further stabilize learning and suppress long-term error growth. This hybrid design enables PENCO to preserve governing physics while mitigating long-term error growth. Through extensive three-dimensional phase-field benchmarks on periodic cubic domains, covering phase ordering, crystallization, epitaxial growth, and complex pattern formation, PENCO demonstrates superior accuracy, stability, and data efficiency compared to state-of-the-art neural operators, including FNO-4D and MHNO, while maintaining physically consistent evolution. The associated dataset and implementation are available at github.com/MBamdad/PENCO.

* Corresponding author. Email: mostafa.bamdad@uni-weimar.de

† Corresponding author. Email: eshaghi.khanghah@iop.uni-hannover.de

‡ Corresponding author. Email: timon.rabczuk@uni-weimar.de

## 1 Introduction

Partial differential equations (PDEs) [1] provide the mathematical foundation for modeling systems that evolve in space and time, capturing complex dynamics in disciplines like fluid flow [2], heat transfer [3], and structural mechanics [4]. A particularly important class of these models is three-dimensional phase-field equations [5], which include formulations for Allen Cahn (AC), Cahn Hilliard (CH), Swift Hohenberg (SH), Phase Field Crystal (PFC), and Molecular Beam Epitaxy (MBE) growth [6, 7]. These equations are powerful tools for simulating interfacial dynamics and material evolution. By using smoothly varying order parameters to represent distinct material phases, they are used to handle problems with complex topological changes, such as crack evolution, boiling, and multiphase flows [6, 8-10]. Consequently, they have become indispensable in studying phenomena such as solidification [11], microstructure evolution [12], and phase separation [13].

Traditional numerical solvers such as finite difference [14] and finite element methods [15, 16] have been extensively used to solve these systems. However, their computational cost can be prohibitive for large-scale or repeated simulations required in design optimization, uncertainty quantification, and inverse analysis [17].

In recent years, machine learning has emerged as a transformative paradigm for accelerating the solution of PDE-governed systems. These approaches can be broadly categorized into three classes. The first are data-driven models, such as NOs [18-24], which learn mappings between function spaces directly from simulation data. The second are pure physics-driven methods, notably Physics-Informed Neural Networks (PINNs) [25-29], which embed governing physical laws into the loss function. The third are hybrid frameworks[30-34], which combine the generalization capability of NOs with physical laws enforced through PDE.

Purely data-driven approaches such as the FNO [35], DeepONet [22, 36], and Graph Neural Operators [37] have achieved remarkable success in modeling complex systems, including phase-field dynamics [38]. Nevertheless, these architectures often require large, high-fidelity datasets, lack embedded physical constraints, and may suffer from overfitting and poor extrapolation [39-41]. Conversely, PINNs mitigate data dependency by incorporating PDE residuals into their loss function [29], enabling physically consistent predictions. This principle has been successfully applied to challenging problems in solid mechanics, such as the analysis of functionally graded porous beams [42], demonstrating the potential of physics-informed learning in engineering contexts. Despite these advantages, standard PINNs struggle with the stiff, multiscale nature of phase-field equations [43] and exhibit limited generalization across varying initial conditions [44], frequently requiring costly retraining [45-47].

The challenges posed by stiffness in PDE learning have been widely investigated, particularly in the context of physics informed solution networks, where stiffness often manifests as optimization imbalance and spectral bias during training [48, 49]. While existing approaches mitigate these effects through architectural modifications or adaptive training strategies, stiffness remains especially challenging in spatiotemporal settings that require long horizon predictions. In such cases, stiffness leads not only to training difficulties but also to accumulated temporal error and numerical instability during rollout. This distinction motivates the need for operator learning frameworks that explicitly incorporate numerical stability and physical structure during temporal evolution.

To address these challenges, hybrid methods have emerged as a promising direction, combining the ability of neural operators to learn mappings between functions with physics-based constraints inspired by PINN formulations. For instance, the PINO enhances an FNO by incorporating PDE residuals into the loss function, enabling it to achieve high accuracy with fewer data points [30]. This area is evolving rapidly, and several strategies exist for integrating physics. Some approaches use adaptive weighting between data and physics losses [50]. Others, such as the variational physics-informed neural operator (VINO), minimize an energy-based variational form, which can reduce the need for high-order derivatives and improve efficiency on fine grids [51]. Related directions have also explored operator-learning formulations based on iterative numerical schemes or hybrid Gaussian process models, offering strong theoretical guarantees or reduced reliance on data-driven losses [52, 53]. However, these approaches are typically designed for specific problem classes or static mappings and do not directly address long-horizon stability and accumulated temporal error in stiff spatiotemporal systems.

While these developments mark significant progress, most operator-learning frameworks still face difficulties in modeling stiff spatiotemporal systems with long temporal horizons. Architectures such as FNO-3D [36] and Phase-Field DeepONet [31] typically employ autoregressive strategies, predicting one or a few time steps at a time. This approach introduces cumulative error and instability in long-time predictions. The FNO-4D partially alleviates this by treating time as an additional input channel, enabling simultaneous learning of spatiotemporal correlations. However, its design lacks explicit causal coupling between consecutive time steps, which is fundamental for physical consistency.

The MHNO [54] was introduced to address the challenges of autoregressive error accumulation and non-causal temporal prediction. Its projection-based temporal mechanism enforces causality while enabling the prediction of all time steps in a single forward pass, resulting in improved temporal coherence and numerical stability compared to purely recurrent or fully parallel designs. However, despite these architectural advantages, MHNO remains fundamentally data-driven and exhibits a strong dependence on large, high-quality datasets for accurate generalization. In data-limited regimes, its prediction error tends to accumulate over time, and the model shows degraded accuracy for unseen initial conditions due to the absence of embedded physical constraints. Consequently, while MHNO enhances temporal connectivity, it does not inherently ensure physical consistency or data efficiency.

Building on these advances, the present work introduces PENCO (Physics-Energy-Numerics-Consistent Operator), a new hybrid framework that combines data efficiency, physical fidelity, and numerical stability within a single operator-learning model. Instead of simply augmenting existing neural operators with residual penalties, PENCO is designed around a principled integration of how physics and numerics interact during temporal evolution. The learning objective mirrors semi-implicit solvers. It balances explicit nonlinear updates with spectrally preconditioned linear components, transferring classical stability into a trainable operator. The model learns from both data and physics simultaneously, guided by collocation-based constraints that maintain temporal coherence and by energy-based regularization that enforces thermodynamic admissibility.

In addition, PENCO stabilizes large-scale structures through a spectral anchoring mechanism that preserves low-wavenumber dynamics while allowing fine-scale features to adaptively evolve. This hybrid formulation captures the essential physical invariants of phase-field systems and suppresses the error accumulation that typically plagues long-horizon operator learning. As a result, PENCO generalizes across different physical regimes and maintains numerical consistency comparable to advanced semi-implicit solvers, bridging the gap between traditional computation and modern data-driven modeling.

The remainder of this study is organized as follows: Section 2 details the methodologies, providing an overview of the FNO-4D, MHNO, and the proposed PENCO frameworks. Section 3 details the data generation procedure and training–budget protocol adopted to assess model performance under data-scarce conditions. Section 4 presents the computational experiments on several 3D phase field equations, including the AC, CH, SH, PFC, and MBE models. Finally, Section 5 concludes the paper with a summary of the findings and potential directions for future research.

## 2 Methodology

This section presents the operator-learning frameworks used to model three-dimensional phase-field dynamics. We first summarize the baseline architectures used for comparison, namely the FNO-4D adapted to four-dimensional spatiotemporal inputs and the MHNO. We then introduce PENCO, which builds upon a spectral neural-operator backbone but fundamentally changes the training objective by embedding physical laws and numerical structure into the learning process. All models are trained to approximate the solution operator of phase-field-type PDEs of the form $\partial_t u = \mathcal{D}(u, \nabla u, \nabla^2 u, \dots)$, where $\mathcal{D}$ denotes the full spatial differential operator. The operator maps an initial to the full spatiotemporal evolution of the field.

### 2.1 Operator-Learning Setting

Let $\Omega \in R^3$ denotes a periodic spatial domain, $t \in [0, T]$ the time interval, and

$$u(\boldsymbol{x}, t) \colon \Omega \times [0, T] \to \mathbb{R} \tag{1}$$

a scalar order parameter governed by a phase-field equation. In an operator-learning setting, we seek to approximate the solution operator

$$\mathcal{G}_\theta \colon\ a(\boldsymbol{x}) \longmapsto u(\boldsymbol{x}, t), \tag{2}$$

Where $a(\boldsymbol{x})$ denotes the initial condition (and possibly additional input channels encoding parameters). A neural operator $\mathcal{G}_\theta$ with trainable parameters $\theta$ is trained from pairs $\{a_j, u_j\}_{j=1}^{N}$, where each $a_j$ is a randomly sampled input function (drawn from a probability distribution $\mu$ ), and $u_j$ is its corresponding solution, generated by a high-fidelity spectral solver (see Appendix). Depending on the architecture, $\mathcal{G}_\theta$ outputs either the entire time sequence in a single forward pass or predicts one step at a time in an autoregressive manner. In what follows, we briefly review the two baseline architectures and then describe the proposed PENCO framework in detail.

## 2.2 Four-Dimensional Fourier Neural Operator (FNO-4D)

The FNO [37, 55] learns mappings between function spaces using spectral convolutions. In the classical formulation, a neural operator

$$\mathcal{G}_\theta : \mathcal{A} \rightarrow \mathcal{U} \tag{3}$$

is represented as a composition of a lifting operator, a sequence of spectral layers, and a projection operator. Given an input function $a(\boldsymbol{x})$, the network first lifts it into a latent representation

$$v_0(\boldsymbol{x}, t) = \mathcal{P}(a(\boldsymbol{x}, \boldsymbol{t})), \tag{4}$$

where $\mathcal{P}$ is a shallow fully connected network acting channel-wise. Each subsequent layer $\tau$ performs a combination of local and global operations:

$$v_{\tau+1}(\boldsymbol{x}, t) \coloneqq \sigma\left(\mathcal{W}_\tau v_\tau(\boldsymbol{x}, t) + \mathcal{F}^{-1}\left(R_\tau \circ \mathcal{F}(v_\tau)(\boldsymbol{k}, t)\right)\right), \qquad \forall \boldsymbol{x} \in \Omega, t \in [0, T] \tag{5}$$

where $\mathcal{W}_\tau$ is a pointwise linear map, $\mathcal{F}$ and $\mathcal{F}^{-1}$ denote the forward and inverse spatial Fourier transforms, $R_\tau$ is a learnable spectral multiplier defined in the wavenumber domain $\boldsymbol{k}$, and $\sigma$ is a nonlinear activation function. In the FNO-4D variant used here, the Fourier transform acts only on the spatial coordinates $\boldsymbol{x} = (x, y, z)$, while the wavenumber vector $\boldsymbol{k} = (k_x, k_y, k_z)$, represents the corresponding spectral modes. Time is treated as an additional input channel rather than a transformed dimension, allowing the model to process multiple time frames in parallel while still leveraging global spatial mixing through spectral convolutions.

This architecture can be expressed compactly as the compositional mapping

$$\mathcal{G}_\theta(a)(\boldsymbol{x}, t) \coloneqq \mathcal{Q} \circ \sigma\left[\underbrace{((\mathcal{W}_\tau + \mathcal{K}_\tau) \circ \cdots \circ (\mathcal{W}_1 + \mathcal{K}_1))}_{\mathcal{T}\ iterative\ layers}\right] \circ \mathcal{P}(a(\boldsymbol{x}, t)), \tag{6}$$

where $\mathcal{Q}$ denotes the final projection layer, $\mathcal{W}_\tau$ are local linear transformations, and $\mathcal{K}_\tau$ represent nonlocal spectral convolutions. Each $\mathcal{K}_\tau$ operates by transforming the latent feature map into Fourier space, scaling its spectral coefficients with a learnable complex-valued filter $R_\tau(\boldsymbol{k})$, and then projecting it back to the spatial domain. This formulation allows the network to learn global interactions through spectral mixing, while $\mathcal{W}_\tau$ capture localized spatial dependencies. The trainable parameters $\theta$ are optimized to minimize the data mean-squared error

$$\min_\theta \frac{1}{N} \sum_{j=1}^{N} \left\| \mathcal{G}_\theta(a_j) - u_j \right\|_2^2, \tag{7}$$

ensuring that the learned operator approximates the true mapping between initial conditions and their corresponding spatiotemporal true solutions $u_j$.

Unlike autoregressive FNO-3D models, which propagate the solution step by step, FNO-4D takes a block of input time slices and predicts the corresponding block of future states in a single forward pass. This design eliminates hidden-state recurrence and reduces explicit error accumulation, but it does not enforce a causal structure between successive time steps at the architectural level. In this work, FNO-4D serves as a strong data-driven baseline that captures rich spatiotemporal correlations but does not incorporate explicit physical or numerical constraints (Figure 2-1).

## 2.3 Multi-Head Neural Operators (MHNO)

The MHNO [54] was proposed to address the complementary limitations of autoregressive error accumulation and non-causal parallel prediction. MHNO retains a spectral neural-operator backbone for spatial processing but introduces a more expressive temporal structure via time-step-specific heads.

First, the initial condition $a(\boldsymbol{x})$ is passed through a shared spectral encoder (similar in spirit to FNO) to produce a high-dimensional latent representation

$$v_a(\boldsymbol{x}) \coloneqq (\mathcal{W}_{\mathcal{L}} + \mathcal{K}_{\mathcal{L}}) \circ \cdots \circ (\mathcal{W}_1 + \mathcal{K}_1) \circ \mathcal{P}(a(\boldsymbol{x})) \tag{8}$$

where $\mathcal{W}_{\mathcal{L}}$ is Fourier integral operators and $\mathcal{K}_{\mathcal{L}}$ is local linear transforms. This encoder is reused for all time steps, providing a shared spatial feature space. At its core, MHNO replaces the static projection operator $\mathcal{Q}$ with a combination of a shared spatial feature extractor and two dynamic, time-step-specific components: projection networks $\{\mathcal{Q}_n\}_{n=1}^{N_t}$ and temporal transition networks $\{\mathcal{H}_n\}_{n=2}^{N_t}$. This dual-component design explicitly couples adjacent time steps, enabling dynamic feature extraction and information passing across the temporal domain. The resulting operator can be written schematically as

$$\mathcal{G}_\theta(a)(\boldsymbol{x}, t_n) \coloneqq \begin{cases} \mathcal{Q}_n\big(v_a(\boldsymbol{x})\big), & \text{for n} = 1, \\ \mathcal{Q}_n\big(v_a(\boldsymbol{x})\big) + \mathcal{H}_n\big(\mathcal{G}_\theta(a)(\boldsymbol{x}, t_{n-1})\big), & \text{for n} > 1. \end{cases} \tag{9}$$

This design allows MHNO to predict all time steps in a single forward pass, while maintaining explicit temporal coupling through $\mathcal{H}_n$. In practice, MHNO improves temporal coherence and stability relative to purely recurrent or fully parallel architectures.

The introduction of time-step-specific projection networks $\mathcal{Q}_n$ and transition operators $\mathcal{H}_n$ gives MHNO a hierarchical interpretation: the shared spectral encoder captures global spatial correlations, while the temporal components act as dynamic filters that adaptively refine features at each time level. This enables the network to represent complex, non-stationary dynamics and long-range temporal dependencies without the need for recurrent hidden states. Each temporal head $\mathcal{Q}_n$ effectively learns the mapping from the encoded spatial field to the corresponding solution snapshot, and the transition networks $\mathcal{H}_n$ propagate information across successive time frames, ensuring causal continuity. Together, these mechanisms allow MHNO to model intricate spatiotemporal behaviors, such as phase evolution, interface migration, and pattern coarsening, within a unified operator-learning framework (Figure 2-1).

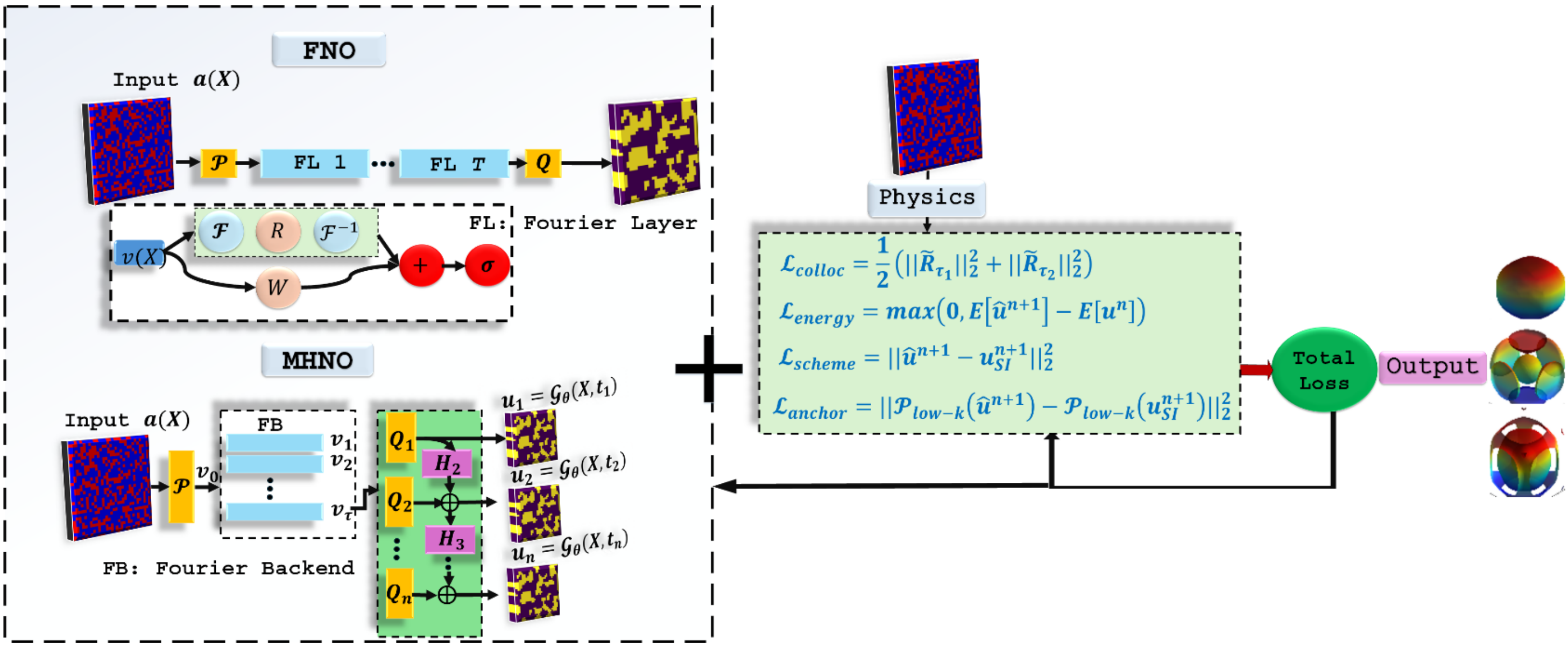

Figure 2-1: Overview of the PENCO training framework applied to either an FNO or an MHNO backbone. PENCO defines the physics-guided training objective while the backbone architectures remain independent.

## 2.4 Physics-Guided Loss Design in PENCO

The PENCO framework embeds physical laws into operator learning by constraining not only the governing equations but also the temporal structure through which solutions evolve. Unlike existing physics-informed operator-learning approaches such as PINO or VINO, which incorporate physics primarily through soft residual or variational penalties evaluated at arbitrary collocation points, PENCO enforces physics guidance through consistency with a stable numerical time-integration scheme.

By aligning operator learning with semi-implicit IMEX time integration and using temporally structured collocation derived from time-integration theory, PENCO constrains the learned dynamics to follow a stable and physically meaningful temporal evolution rather than relying on pointwise residual minimization. In this sense, numerical-scheme consistency constitutes a qualitatively different form of physics guidance. The resulting training objective combines data supervision with IMEX-aligned temporal constraints, energy dissipation, spectral stability, and collocation-based enforcement of the governing dynamics, directly targeting stability and error accumulation in long-horizon simulations of stiff phase-field systems.

To train the neural operator $\mathcal{G}_\theta$, PENCO combines a standard data-fitting term with a physics-guided regularization term. The total objective is expressed as

$$\mathcal{L}_{total} = (1-\lambda)\alpha\mathcal{L}_{data} + \lambda\mathcal{L}_{phys}, \quad \lambda = [0,1], \tag{10}$$

where the coefficient $\alpha$ provides a scaling factor that balances the magnitudes of the data and physics terms, ensuring that neither dominates the optimization. In practice, $\alpha$ is chosen such that the two contributions remain of comparable order during training, allowing the model to learn consistently from both empirical trajectories and physical constraints. For all case studies in this work $\alpha$ is set either $10^2$ $or$ $10^3$, depending on the characteristic scales of the governing equation, ensuring that the data and physics contributions remain comparable throughout training. The data loss measures direct prediction error against ground truth,

$$\mathcal{L}_{data} = ||\hat{u}^{n+1} - u_{\text{true}}^{n+1}||_2^2 \tag{11}$$

and $\mathcal{L}_{phys}$ represents the physics-informed component that constrains the model to obey the governing dynamics.

The physics-based regularization term integrates four complementary components, each addressing a different aspect of physical or numerical fidelity,

$$\mathcal{L}_{phys} = w_1(\mathcal{L}_{colloc} + w_2\mathcal{L}_{anchor}) + w_3\mathcal{L}_{scheme} + w_4\mathcal{L}_{energy} \tag{12}$$

where $\mathcal{L}_{colloc}$ enforces the differential dynamics by evaluating the PDE residual at Gauss–Lobatto points. This penalizes any mismatch between the predicted field and the local PDE behavior, ensuring that the network does not drift into non-physical trajectories even when the data term is weak or noisy.

$\mathcal{L}_{scheme}$ promotes temporal consistency by comparing each predicted update with a stable reference evolution of the PDE. Instead of using a fully explicit step, which would become unstable for stiff phase field models, the reference update applies the linear smoothing operator in Fourier space together with the nonlinear term. This produces a well-conditioned target that reflects the natural interaction of diffusion and nonlinearity and provides a reliable guide for the network during training.

$\mathcal{L}_{energy}$ enforces thermodynamic admissibility by penalizing any increase in the free energy, ensuring that the predicted evolution follows the natural energy-dissipating behavior of phase-field systems and does not produce non-physical growth of energetic modes.

The low-frequency ($\mathcal{L}_{anchor}$) anchor term penalizes discrepancies between the large-scale Fourier modes of the predicted field and those of the physics-based semi-implicit update. This prevents slow spectral drift or unphysical growth of large-scale structures during training which is a common failure mode of neural operators.

The epoch-dependent weights $w_2$ and $w_3$ evolve smoothly during training as

$$w_2(e) = 0.25 + 0.6\left(\frac{e}{E-1}\right)^2, \qquad w_3(e) = 0.32 - 0.12\frac{e}{E-1} \tag{13}$$

where $e$ denotes the current epoch and $E$ the total number of epochs. In all experiments, we fix $w_1 = 10^{-3}$ for the collocation–anchor block and $w_4 = 0.3$ for the energy term, which keeps their scales consistent with the scheme loss and prevents any component from dominating the optimization. The epoch dependent weights $w_2(e)$ and $w_3(e)$ then adjust the remaining contributions during training. At the beginning of training the scheme consistency term plays a stronger role and helps the model settle into stable short horizon updates. As training progresses, the increasing influence of the anchor term reinforces long range spectral stability and suppresses low frequency drift. This gradual shift in emphasis ensures smooth convergence and maintains a balanced interaction among all physics-based components across the different phase-field models.

Together, these components guide the network to capture both the local PDE behavior and the global energetic trends characteristic of phase-field systems, achieving stability, accuracy, and generalization even in data-scarce regimes.

### 2.4.1 $L^2$ Gauss–Lobatto Collocation Residual (P)

To ensure that the predicted field evolution satisfies the governing partial differential equation not only at discrete time levels but also within each time step, PENCO employs an $L^2$ Gauss–Lobatto collocation residual. Let $u(\boldsymbol{x}, t)$ be a scalar field on a periodic cube $\Omega = [-L/2, L/2]^3$ evolving by a phase-field-type PDE

$$\partial_t u = \mathcal{D}(u), \tag{14}$$

the residual over a time step $[t_n, t_{n+1}]$ with increment $\Delta t$ is defined as

$$R_\tau = \frac{\hat{u}^{n+1} - u^n}{\Delta t} - \mathcal{D}((1-\tau)\, u^n + \tau\, \hat{u}^{n+1}), \tag{15}$$

where $R_\tau$ is the PDE's residual at Gauss–Lobatto collocation points, and $\tau \in [0,1]$ denotes the normalized collocation location. With a uniform time step $\Delta t$, we denote $u^n(\boldsymbol{x}) \approx u(\boldsymbol{x}, t_n)$, $t_{n+1} = t_n + \Delta t$, and the neural operator $\mathcal{G}_\theta$ predicts the next state $\hat{u}^{n+1} = \mathcal{G}_\theta(u^n)$. PENCO adopts two symmetric Gauss–Lobatto nodes,

$$\tau_{1,2} \in \left\{\frac{1}{2} \pm \frac{1}{2\sqrt{5}}\right\}, \tag{16}$$

which provide balanced sampling around the temporal midpoint and enhance stability during training. To improve numerical robustness, the residuals are computed in a normalized form,

$$\widetilde{\mathrm{R}}_\tau = \frac{\dot{u}^n}{||\dot{u}^n||_2} - \frac{\mathcal{D}(u_\tau)}{||\mathcal{D}(u_\tau)||_2}, \tag{17}$$

where $\dot{u}^n = (\hat{u}^{n+1} - u^n)/\Delta t$ and $u_\tau = (1-\tau)\, u^n + \tau\, \hat{u}^{n+1}$. This normalization stabilizes gradient magnitudes during early training and balances the relative influence of stiff linear and nonlinear components, leading to smoother convergence. The associated loss term is then expressed as $\mathcal{L}_{\text{colloc}}$

$$\mathcal{L}_{\text{colloc}} = \frac{1}{2}\left(||\widetilde{\mathrm{R}}_{\tau_1}||_2^2 + ||\widetilde{\mathrm{R}}_{\tau_2|}|_2^2\right), \tag{18}$$

The choice of a two-point Gauss–Lobatto collocation around the temporal midpoint is motivated by both stability considerations and empirical evaluation. We tested several alternative strategies, including midpoint-only collocation as well as three-point and five-point Gauss–Legendre collocation. While these approaches achieved comparable accuracy for simpler phase-field models such as the AC equation, their performance degraded for higher-order and more complex systems, including the CH, SH, PFC, and MBE equations, where accumulated temporal error and instability became more pronounced. In contrast, the symmetric two-point Gauss–Lobatto collocation consistently provided robust accuracy and stability across all tested models, offering an effective balance between temporal accuracy, numerical robustness, and computational efficiency. For this reason, it is adopted throughout this work.

The collocation residual therefore provides a strong constraint on the local governing dynamics within each time step. To ensure that this local accuracy translates into stable long-horizon evolution for stiff phase-field systems, PENCO augments the collocation term with additional physics-guided losses that preserve thermodynamic consistency, enforce numerical stability, and control large-scale spectral behavior during training.

### 2.4.2 Energy consistency (E)

Most phase-field models arise as gradient flows of a free-energy functional, and their physical solutions satisfy a monotonic decay of the total free energy over time. If $E[u]$ denotes the free energy of the state $u(\boldsymbol{x}, t)$, then the thermodynamic structure of these models requires $E[u^{n+1}] \leq E[u^n]$ at every discrete time step, ensuring that the system relaxes toward equilibrium rather than accumulating artificial energy. In general, the free energy consists of a bulk contribution together with gradient-dependent regularization terms, which can be written in the generic form

$$E[u] = \int_\Omega \left( F(u) + \frac{1}{2}\epsilon^2 |\nabla u|^2 + H(u, \nabla u, \Delta u, \dots) \right) d\boldsymbol{x}, \tag{19}$$

where $F(u)$ denotes the local thermodynamic potential (often a double-well form), the gradient term $\frac{1}{2}\epsilon^2 |\nabla u|^2$ penalizes sharp interfaces and introduces the characteristic interface width $\epsilon$, and the functional $H(.)$ accounts for higher-order spatial derivatives that appear in extended phase-field models such as PFC and MBE.

While this expression is presented in a generic form for clarity, each benchmark equation considered in this work employs its corresponding model-specific energy functional. For lower-order models such as AC and CH, the energy includes only the bulk potential and first-order gradient regularization. For higher-order models, including SH, PFC, and MBE, additional Laplacian or higher-derivative contributions are explicitly incorporated. All spatial derivatives appearing in the energy are evaluated spectrally in Fourier space, consistent with the periodic setting and the numerical discretization adopted throughout this study.

Conventional numerical schemes enforce energy dissipation by construction, whereas a purely data-driven neural operator may violate this property, leading to spurious energy growth and unstable long-horizon predictions. To prevent such behavior, PENCO introduces a one-sided energy dissipation penalty,

$$\mathcal{L}_{energy} = \max(0,\, E[\hat{u}^{n+1}] - E[u^n]), \tag{20}$$

which is activated only when the predicted update increases the free energy. This constraint enforces thermodynamic admissibility during training and improves long-term stability, complementing the local PDE residual and numerical scheme consistency constraints.

### 2.4.3 Numerical (scheme) Consistency (N)

In addition to enforcing PDE residuals directly, PENCO further improves stability by aligning the network's predictions with a semi-implicit time-stepping scheme that is widely used in numerical simulations of phase-field models. Such schemes are based on the implicit–explicit (IMEX) principle, where stiff linear terms $L(u)$ are treated implicitly to enhance stability, while nonlinear terms $\mathcal{N}(u)$ are evaluated explicitly to avoid solving nonlinear systems at each step. Let the general phase-field PDE be written in operator-split form as

$$\frac{u^{n+1} - u^n}{\Delta t} = -\mathcal{M}\big(L(u^{n+1}) + \mathcal{N}(u^n)\big), \tag{21}$$

where $\mathcal{M}$ is the mobility operator that defines the geometry of the gradient flow (for instance, $\mathcal{M} = I$ in AC, and $\mathcal{M} = -\Delta$ in CH). The above relation can be rearranged into the linear system

$$(I + \Delta t\, \mathcal{M}L)u^{n+1} = u^n - \Delta t\, \mathcal{M}\mathcal{N}(u^n), \tag{22}$$

solving this expression yields the semi-implicit ($SI$) numerical update

$$u_{SI}^{n+1} = (I + \Delta t\, \mathcal{M}L)^{-1}\big(u^n - \Delta t\, \mathcal{M}\mathcal{N}(u^n)\big), \tag{23}$$

which we refer to as the numerical teacher. This $u_{SI}^{n+1}$ is not a learned quantity but the deterministic update produced by the IMEX scheme when starting from state $u^n$. To encourage the neural operator to remain consistent with this stable evolution, PENCO introduces the scheme consistency loss

$$\mathcal{L}_{scheme} = ||\hat{u}^{n+1} - u_{SI}^{n+1}||_2^2, \tag{24}$$

which penalizes deviations from the IMEX update. This consistency term serves a dual purpose. First, by aligning the model predictions with a semi-implicit IMEX update, it transfers the inherent stability of numerical solvers into the learning process and prevents the uncontrolled accumulation of errors over long horizons. Second, it guides the neural operator to respect the natural linear–nonlinear splitting structure of the governing PDE, which enhances its ability to generalize beyond the specific training data.

To further reinforce stability at large spatial scales, the scheme consistency is extended into the spectral domain through a low-frequency anchoring constraint, which minimizes deviations between the coarse Fourier components of the prediction and those of the semi-implicit reference:

$$\mathcal{L}_{anchor} = ||\mathcal{P}_{low-k}(\hat{u}^{n+1}) - \mathcal{P}_{low-k}(u_{SI}^{n+1})||_2^2, \tag{25}$$

where $\mathcal{P}_{low-k}(.)$ denotes projection onto the lowest-wavenumber fraction of the spectrum. This spectral anchoring prevents phase drift and unphysical growth of long-wavelength modes, thereby preserving coherent large-scale structures throughout training. Together, the real-space and spectral components of numerical consistency provide PENCO with strong stability and accuracy across extended temporal horizons.

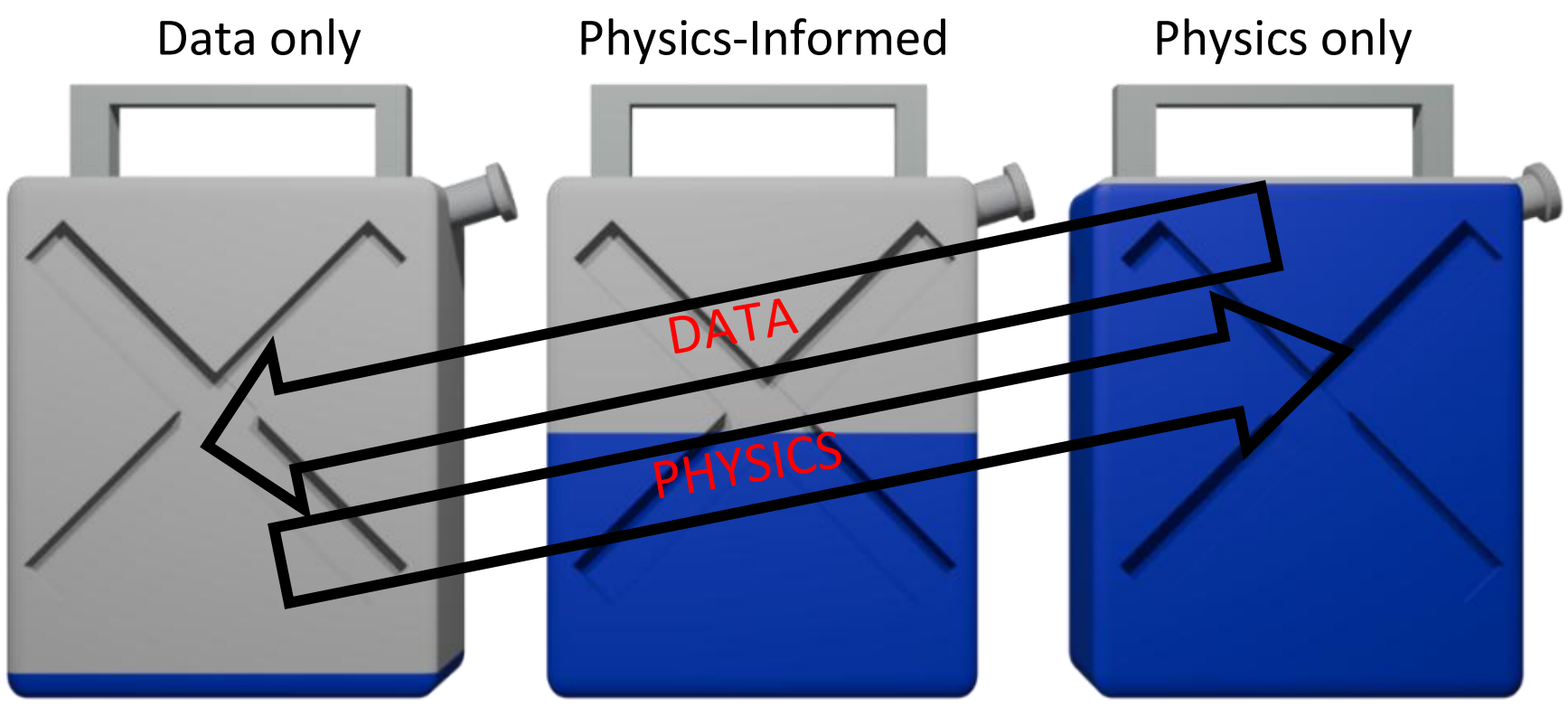

Figure 2-2: A conceptual illustration of training paradigms. PENCO employs the hybrid strategy (center), where the model is filled with knowledge from both data and physics to ensure physical consistency.

## 2.5 Summary of the PENCO Training Pipeline

This algorithm summarizes a single PENCO optimization step, where supervised data fitting is jointly combined with collocation residual enforcement, semi-implicit scheme consistency, low-frequency spectral anchoring, and energy dissipation constraints.

I. Sample batch of states $\{u^n, u_{\text{true}}^{n+1}\}$.

II. Predict $\hat{u}^{n+1} = \mathcal{G}_\theta\,(u^n$.

III. Compute $\mathcal{L}_{data} = ||\hat{u}^{n+1} - u_{\text{true}}^{n+1}||_2^2$.

IV. Compute the normalized Gauss–Lobatto residuals $\widetilde{\text{R}}_{\tau_1}, \widetilde{\text{R}}_{\tau_2} \rightarrow \mathcal{L}_{\text{colloc}}$.

V. Compute IMEX teacher $u_{SI}^{n+1} \rightarrow \mathcal{L}_{scheme}$.
VI. Low-k projection $\rightarrow \mathcal{L}_{anchor}$.
VII. Compute energy $E[.] \rightarrow \mathcal{L}_{energy}$.
VIII. Total loss assembly $\mathcal{L}_{total}$.
IX. Update $\theta$ by backpropagation.

# 3 Data Generation and Training–Budget Protocol

## 3.1 Data – Generation

Unlike purely data-driven neural operators, which heavily depend on large simulation corpora to achieve acceptable accuracy, PENCO is designed to operate effectively even when only a very limited number of trajectories are available. In our experiments, the training regime intentionally reflects this data-scarce setting, with only $\mathrm{N} = \{50, 100, 200\}$ trajectories provided for supervision. These datasets serve merely as auxiliary physical examples, while the governing dynamics are primarily learned through the physics-based loss components.

The reference trajectories are obtained by numerically integrating five benchmark three-dimensional phase-field models (AC, CH, SH, PFC, and MBE) on periodic cubic domains. The simulations are carried out using a Fourier-spectral discretization [7] in space combined with a semi-implicit time integrator, a formulation that provides high numerical fidelity and excellent stability for long-horizon phase-field dynamics. The detailed mathematical formulation and implementation of this spectral solver are provided in Appendix A. Each trajectory is initialized from a randomized Gaussian Random Field whose statistical parameters are resampled for every simulation, producing a diverse ensemble of physically plausible morphologies without requiring any handcrafted or engineered initial seeds. This stochastic initialization mechanism naturally spans a wide distribution of interface geometries, characteristic wavelengths, and phase-separation patterns, thereby supplying minimal but sufficiently rich supervision to guide the operator during training.

A key objective in our data generation strategy was to create a rich and diverse set of initial conditions to ensure our trained models are robust and can generalize effectively. To achieve this, we employed a two-pronged approach. The primary method involved initializing simulations with Gaussian Random Fields (GRFs). Crucially, rather than using fixed statistical parameters, we randomized the GRF's controlling parameters and the subsequent thresholding value for each simulation. This process generated a wide spectrum of initial random morphologies with varying characteristic length scales and volume fractions.

All numerical data generation and training are performed using PyTorch on an NVIDIA A100 GPU (40GB memory), ensuring consistent floating-point conventions between the spectral solver and neural operator training. All reported experiments are conducted on $32^3$ spatial grids. While the proposed framework is not inherently restricted to this resolution, increasing the grid size (e.g., $128^3$ or $256^3$) in three-dimensional phase-field simulations substantially increases the size of the generated spectral datasets, leading to higher memory requirements for storage and additional data-loading overhead during training. The chosen resolution therefore enables stable and efficient execution of both data generation and learning within the available computational resources.

## 3.2 Training–Budget Protocol

A central goal of this study is to assess neural operator performance in a data-scarce regime, where only a limited number of reference trajectories are available for supervision. Accordingly, we consider training sets of size $\mathrm{N} = \{50, 100, 200\}$, which are one to two orders of magnitude smaller than those typically required by purely data-driven neural operators (which often rely on $\sim 2000 - 5000$ trajectories to achieve stable accuracy). To enable a fair comparison across different dataset sizes and across learning regimes (data-driven, hybrid, and pure-physics), we enforce two safeguards:

### 3.2.1 Fixed evaluation set

The held-out test set consists of a constant number of trajectories (50 for every PDE), ensuring that evaluation is stable and directly comparable across dataset sizes.

### 3.2.2 Fixed optimization budget

The number of gradient updates is kept constant for a given PDE, independent of the number of available training trajectories. This prevents larger datasets from implicitly receiving more optimizer steps and therefore more learning opportunity. In the pure-physics limit ($\lambda = 1$), this guarantee is essential: the model should improve because it better satisfies the PDE, not because it was allowed to take more gradient steps.

A consistent neural architecture is maintained throughout all experiments, employing a compact spectral operator with two layers, moderate latent width, and approximately 800k trainable parameters. The design is balanced for expressivity and stability in 3D spatiotemporal learning on $32^3$ grids. Table 3-1 summarizes our fixed-budget training protocol, where the base steps per epoch (SPE) were determined empirically to balance accuracy against computational requirements. Effective SPE scales with dataset size relative to the reference ($N_{ref} = 50$) as $SPE \times (N/N_{ref})$. Total gradient updates (the count of optimizer weight updates) are computed as effective $SPE_{eff} \times Epoch$ for hybrid training ($\lambda \neq 1$), while pure physics ($\lambda = 1$) uses the fixed base $SPE \times Epoch$ to maintain constant computational budget regardless of dataset size. This computational budgeting strategy ensures that performance differences across $\mathrm{N} = 50, 100, 200$ arise solely from data efficiency rather than hidden computational advantages. Finally, the average epoch time in the last column reflects PENCO training under the MHNO architecture with a hybrid weighting set to $\lambda = 0.25$.

Table 3-1: Training protocol and computational budget across phase-field equations.

| PDEs | N | Batch | SPE | Epoch | $SPE_{eff}$ $\lambda \neq 1$ | Total updates $\lambda = 1$ | Total updates $\lambda \neq 1$ | Avg. epoch time (s) |
|---|---|---|---|---|---|---|---|---|
| AC | 50 | 8 | 10 | 50 | 10 | 500 | 500 | 0.27 |
| | 100 | 8 | 10 | 50 | 20 | 500 | 1000 | 0.63 |
| | 200 | 8 | 10 | 50 | 40 | 500 | 2000 | 1.15 |
| **CH** | 50 | 8 | 30 | 50 | 30 | 4000 | 1500 | 2.1 |
| | 100 | 8 | 30 | 50 | 60 | 4000 | 3000 | 3.9 |
| | 200 | 8 | 30 | 50 | 120 | 4000 | 6000 | 8.1 |
| **SH** | 50 | 8 | 5 | 150 | 5 | 750 | 750 | 0.42 |
| | 100 | 8 | 5 | 150 | 10 | 750 | 1500 | 0.86 |
| | 200 | 8 | 5 | 150 | 20 | 750 | 3000 | 1.45 |
| **PFC** | 50 | 8 | 10 | 50 | 10 | 500 | 500 | 0.79 |
| | 100 | 8 | 10 | 50 | 20 | 500 | 1000 | 1.7 |
| | 200 | 8 | 10 | 50 | 40 | 500 | 2000 | 3.3 |
| **MBE** | 50 | 8 | 30 | 50 | 30 | 1500 | 1500 | 3.71 |
| | 100 | 8 | 30 | 50 | 60 | 1500 | 3000 | 6.71 |
| | 200 | 8 | 30 | 50 | 120 | 1500 | 6000 | 12.13 |

## 4 Numerical Experiments

This section evaluates the performance of the proposed physics-guided neural operator PENCO in comparison with the data-driven frameworks FNO-4D and MHNO across five standard three-dimensional phase-field equations. These benchmarks collectively capture interfacial motion, phase separation, and pattern-forming dynamics, and are unified by an underlying gradient-flow structure, in which the dynamics evolve to dissipate a model-specific free-energy functional (with conserved gradient flows in CH, PFC, and MBE). They can be expressed under the generalized formulation

$$\frac{\partial u}{\partial t} = \alpha_1 \left[\nabla^2 u - \frac{1}{\epsilon^2} f'(u)\right] + \alpha_2 \nabla^2[-\epsilon^2 \nabla^2 u + f'(u)] + \alpha_3[-u^3 - (1-\epsilon)u + 2\nabla^2 u - \nabla^4 u] + \alpha_4 \nabla^2[u^3 + (1-\epsilon)u + 2\nabla^2 u + \nabla^4 u] + \alpha_5\left[-\epsilon\nabla^4 u - \nabla \cdot \left((1-|\nabla u|^2)\nabla u\right)\right] \tag{26}$$

$$u(\boldsymbol{x}, 0) = u_0(\boldsymbol{x}), \qquad \boldsymbol{x} \in [-L/2, L/2], \qquad t \in (0, T]$$

where $u_0$denotes the prescribed initial field. The spatial domain is periodic and defined over $\boldsymbol{x}$, with time $t$. The coefficients $\alpha_1, \dots, \alpha_5$ specify the governing models, AC, CH, SH, PFC, and MBE respectively.

Each equation is simulated using four surrogate models, FNO-4D, MHNO, the proposed hybrid PENCO, and a pure-physics operator, and all predictions are directly compared against high-fidelity numerical reference solutions generated by a semi-implicit Fourier-spectral solver. All models are trained under identical data-scarce regimes and evaluated using the same spatial resolution and time-stepping parameters.

To assess predictive accuracy, all neural operators are evaluated against the high-fidelity numerical solution using the normalized $L_2$- error

$$\text{Error} = \frac{\sqrt{\sum_{i=1}^{N} |\mathcal{U}(x_i, t_i) - u(x_i, t_i)|^2}}{\sqrt{\sum_{i=1}^{N} |u(x_i, t_i)|^2}}, \tag{27}$$

where $u(\boldsymbol{x}_i, t_i)$ denotes the reference field and $\mathcal{U}(\mathrm{x_i}, \mathrm{t_i})$ the model prediction at discrete space–time samples. This metric provides a uniform measure of deviation from the true dynamics and enables consistent quantitative comparison across all PDEs, resolutions, and rollout lengths.

## 4.1 In-Distribution Performance

### 4.1.1 Phase Separation Kinetics: The Allen-Cahn Equation

We first benchmark our proposed models on the 3D AC equation, a fundamental model for phase separation and coarsening dynamics. Its versatility makes it a cornerstone in computational science, with its explicit use in wide-ranging applications such as simulating ductile fracture [56], interface dynamics simulation [57], and crystal growth [58]. Within our unified phase-field formulation (Eq. 25), it corresponds to setting $\alpha_1 = 1$ ($\alpha_{2-5} = 0$), yielding

$$\frac{\partial u(\boldsymbol{x},t)}{\partial t} = \nabla^2 u(\boldsymbol{x},t) - \frac{1}{\epsilon^2} f'\big(u(\boldsymbol{x},t)\big), \qquad (28)$$
$$u(x,0) = u_0(x), \qquad \boldsymbol{x} \in [-L/2, L/2], \qquad t \in (0,T].$$

The system's evolution is governed by the interplay between two terms. The first is the reaction term, derived from the derivative of a double-well energy potential, $f(u) = \frac{1}{4}(u^2 - 1)^2$. Its form, $f'(u) = u^3 - u$, drives the phase-field variable $u$ towards one of the two stable, pure-phase states ($u = \pm 1$). This is counteracted by the diffusion term, $\nabla^2 u$, which acts to minimize the interfacial energy between phases, analogous to surface tension, thereby smoothing complex geometries over time. The equation as a whole, ${\partial u}/{\partial t}$ , describes how these two competing forces evolve the system, with the parameter $\epsilon$ balancing them and controlling the thickness of the final interface.

The primary objective is to train a neural operator, $\mathcal{G}_\theta$, that learns the solution operator mapping an initial condition $u_0(\boldsymbol{x})$ to the full spatiotemporal evolution of the system:

$$\mathcal{G}_\theta\colon\ u(\boldsymbol{x},0) \ \rightarrow\ \{u(\boldsymbol{x},t)\}_{t\in(0,T]} \qquad (29)$$

Each trajectory is defined on a uniform grid of $32^3$ points, the interface width parameter $\epsilon = 0.1$, and the temporal resolution of $N_t = 100$ with uniform time step-size $\Delta t = 0.0001$.

We first assess the model's performance on test data drawn from the same GRF-based distribution used for training. Figure 4-1 and Figure 4-2 report the spatial distribution of the relative pointwise error between the predicted and exact solutions at $t = \{20, 40, 60, 80, 100\}\Delta t$ for FNO-4D, MHNO, and PENCO, compared against the numerical reference. For brevity, the hybrid physics-guided model based on the MHNO architecture is referred to simply as PENCO throughout the plots. Figure 4-1 shows two-dimensional contour slices of the pointwise relative error, while Figure 4-2 presents the corresponding three-dimensional error fields projected onto the exact iso-surface geometry, highlighting where each method deviates from the true dynamics during the coarsening process.

Both data-driven baselines exhibit increasing error accumulation over time. Their deviations become more pronounced in later stages of evolution, with errors concentrating near interfacial regions and complex geometric features. This behavior is most evident in FNO-4D, which shows the largest relative error magnitudes during the later stages of coarsening.

PENCO, trained with a moderate physics weight ($\lambda = 0.25$), consistently yields significantly lower error levels across all time frames. It maintains accurate interfacial structure and remains closely aligned with the reference solution even at the final time. The physics-guided constraints effectively suppress long-horizon drift, which is a persistent failure mode in purely data-driven operators. The pure-physics configuration ($\lambda = 1$) also achieves stable error behavior, confirming the stabilizing role of the PDE structure, although it does not match the uniformly low error attained by the hybrid model.

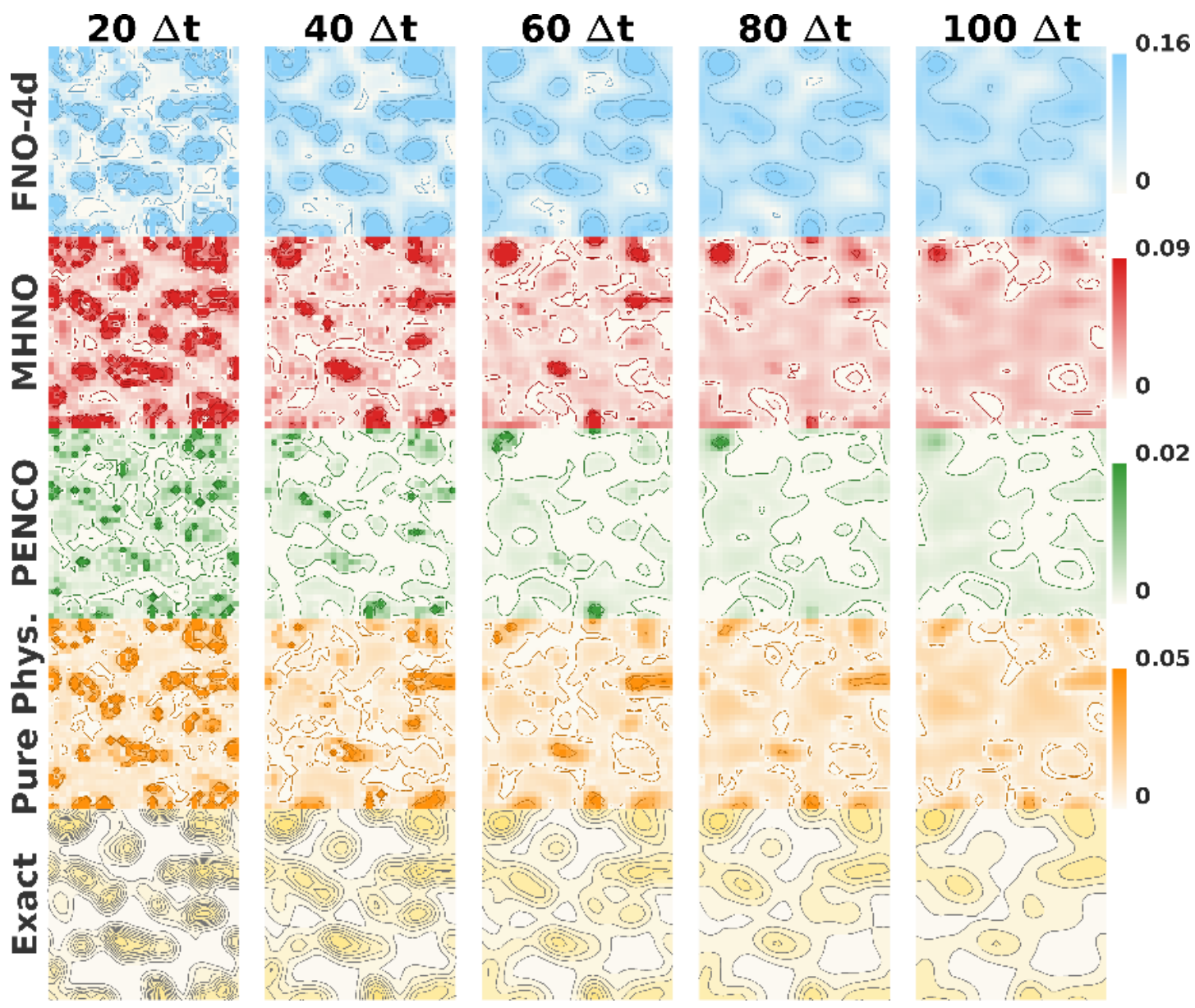

Figure 4-1: *Pointwise relative error maps of the predicted* AC system *compared with its exact reference at different time frames.* ($32^3$ *grid,* $\Delta t = 0.0001$ $\lambda = 0.25$*,* $\epsilon = 0.1$*,* $L = 2$*,* $N = 200$)

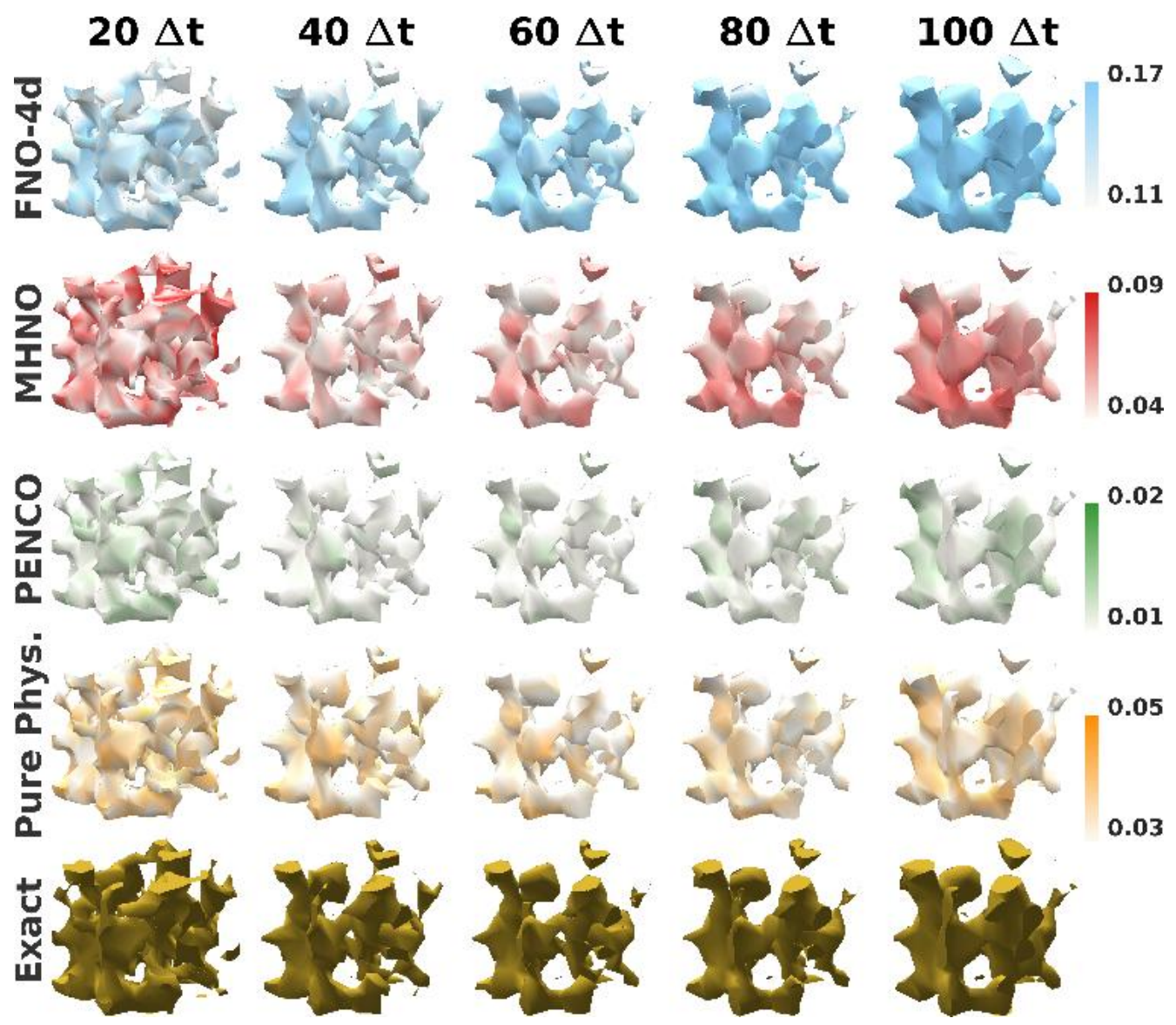

Figure 4-2: Three-dimensional visualization of the pointwise relative error on the AC iso-surface for the same models and simulation parameters as in Figure 4.1.

Figure 4-3 summarizes the convergence behavior and long-horizon accuracy for the AC system using $N = 200$ training samples. The total loss curves clearly distinguish the behavior of the different methods. PENCO converges rapidly and avoids the oscillations observed in the purely data-driven baselines, demonstrating that the physics-guided terms stabilize the optimization process and suppress gradient drift. The pure physics model also converges smoothly, but at a slower rate due to the absence of data supervision.

The relative $L_2$ error curves highlight the effect of physical regularization during rollout. FNO-4D and MHNO exhibit a pronounced error increase immediately after the initial transient, reflecting the tendency of purely data-driven operators to overshoot when the trajectory leaves the training manifold. In contrast, PENCO maintains a low and nearly time-independent error across the full horizon. Both PENCO variants provide substantially lower error than the baselines, with PENCO-MHNO achieving the best overall stability and accuracy.

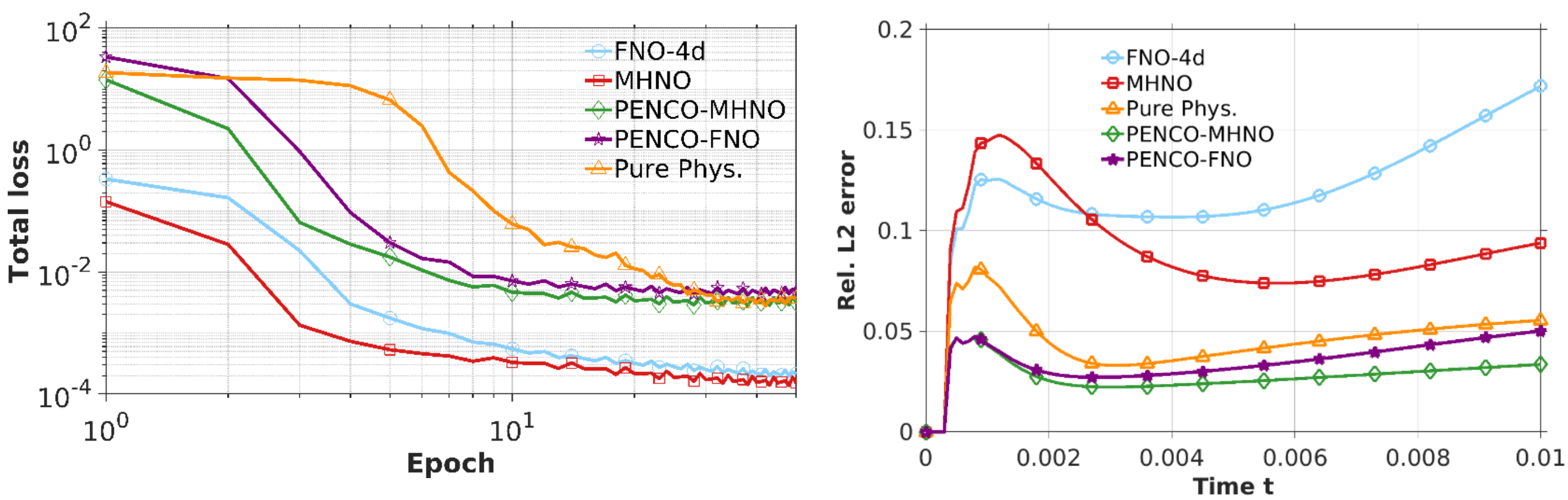

Figure 4-3: Total training loss and relative $L_2$-error for the AC system.

The quantitative comparisons in Table 4-1 show that physics-guided regularization produces a substantial reduction in long-horizon error across all data regimes. In the low-data setting with $N = 50$, the PENCO models (built on the MHNO architecture, as in all following tables) achieve a dramatic gain in accuracy, lowering the final-time error by approximately **80–85%** relative to the data-driven baselines. This corresponds to a five- to six-fold improvement, highlighting the critical role of physical structure when training data are scarce. The pure-physics configuration also performs markedly better than both FNO-4D and MHNO, indicating that the PDE constraints alone already stabilize the dynamics and yield reliable predictions.

As the training set increases to $N = 100$, the advantage of the hybrid formulation remains significant. PENCO delivers an accuracy improvement of roughly **70–80%** compared to the baselines, showing that the integration of physics continues to provide superior long-term fidelity even when more data are available. With $N = 200$, the hybrid operators retain a strong margin, improving accuracy by **about 65–70%** relative to MHNO and nearly **90%** relative to FNO-4D. These consistent gains across all training regimes demonstrate that incorporating the governing physical structure into the learning process substantially enhances both accuracy and stability.

Table 4-1: Relative $L_2$ norm errors of AC equation ($L = 2$, grid $32^3$, $\epsilon = 0.1, \Delta t = 0.0001$).

| N | Method | $0\Delta t$ | $20\Delta t$ | $40\Delta t$ | $60\Delta t$ | $80\Delta t$ | $100\Delta t$ |
|---|---|---|---|---|---|---|---|
| 50 | FNO-4d | 0 | 0.25 | 0.24 | 0.28 | 0.33 | 0.37 |
| 50 | MHNO | 0 | 0.18 | 0.18 | 0.24 | 0.31 | 0.36 |
| 50 | PENCO ($\lambda = 0.25$) | 0 | 0.04 | 0.05 | 0.06 | 0.07 | 0.08 |
| 50 | PENCO ($\lambda = 0.50$) | 0 | 0.05 | 0.05 | 0.06 | 0.07 | 0.08 |
| 50 | PENCO ($\lambda = 0.75$) | 0 | 0.06 | 0.05 | 0.06 | 0.07 | 0.08 |
| 50 | Pure Phys. ($\lambda = 1.0$) | 0 | 0.06 | 0.04 | 0.05 | 0.06 | 0.06 |
| 100 | FNO-4d | 0 | 0.14 | 0.12 | 0.13 | 0.16 | 0.2 |
| 100 | MHNO | 0 | 0.17 | 0.1 | 0.08 | 0.09 | 0.11 |
| 100 | PENCO ($\lambda = 0.25$) | 0 | 0.03 | 0.03 | 0.03 | 0.04 | 0.04 |
| 100 | PENCO ($\lambda = 0.50$) | 0 | 0.03 | 0.03 | 0.03 | 0.03 | 0.04 |
| 100 | PENCO ($\lambda = 0.75$) | 0 | 0.04 | 0.02 | 0.03 | 0.03 | 0.03 |
| 100 | Pure Phys. ($\lambda = 1.0$) | 0 | 0.06 | 0.04 | 0.05 | 0.06 | 0.06 |
| 200 | FNO-4d | 0 | 0.11 | 0.11 | 0.11 | 0.12 | 0.16 |
| 200 | MHNO | 0 | 0.11 | 0.06 | 0.06 | 0.06 | 0.06 |
| 200 | PENCO ($\lambda = 0.25$) | 0 | 0.03 | 0.02 | 0.02 | 0.02 | 0.02 |
| 200 | PENCO ($\lambda = 0.50$) | 0 | 0.03 | 0.02 | 0.02 | 0.02 | 0.02 |
| 200 | PENCO ($\lambda = 0.75$) | 0 | 0.04 | 0.01 | 0.01 | 0.02 | 0.02 |
| 200 | Pure Phys. ($\lambda = 1.0$) | 0 | 0.06 | 0.04 | 0.05 | 0.06 | 0.06 |

### 4.1.2 Conserved Phase Separation: The Cahn-Hilliard Equation

The CH equation models conserved phase separation and introduces a higher level of complexity due to its fourth-order spatial operator. Within the unified framework it is obtained by activating the $\alpha_2$ term, and its dynamics are driven by the Laplacian of the chemical potential, which enforces strict mass conservation and curvature-controlled coarsening. This mechanism underlies a wide range of physical processes, including microstructure evolution in alloys and battery electrodes [59], moving contact lines in confined flows [60], air–water interface dynamics [61], and diffuse-interface tumor growth [62]. The conservation law makes the equation a demanding benchmark for surrogate models, especially when training data are limited.

The CH error visualizations in Figure 4-4 and Figure 4-5 report the spatial distribution of the pointwise relative error between the predicted and reference solutions for FNO-4d, MHNO, PENCO, and the pure-physics variant as time advances. The three baseline models gradually exhibit larger deviations from the numerical solution, with errors concentrating near interfacial regions and small-scale structures. This behavior is also reflected in Figure 4-6, where the relative $L_2$-error of these three models follows a similar increasing trend, with the pure physics model only marginally more accurate than the two data-driven networks. In all three cases, the lack of a strong corrective mechanism allows numerical diffusion and small mass imbalance to accumulate, leading to increasing morphological discrepancies over time.

In contrast, PENCO with $\lambda = 0.25$, built on the MHNO architecture, maintains significantly lower error levels throughout the rollout. In Figure 4-4 and Figure 4-5 its error maps remain consistently weak, indicating sharper interface preservation, improved domain topology, and a coarsening behavior that stays closely aligned with the ground truth. The corresponding curve in Figure 4-6 remains well below all other methods and shows only mild growth over time. This marked improvement confirms that the added physics guidance is essential for stabilizing the conservative dynamics and preventing the long-horizon drift observed in the baseline models.

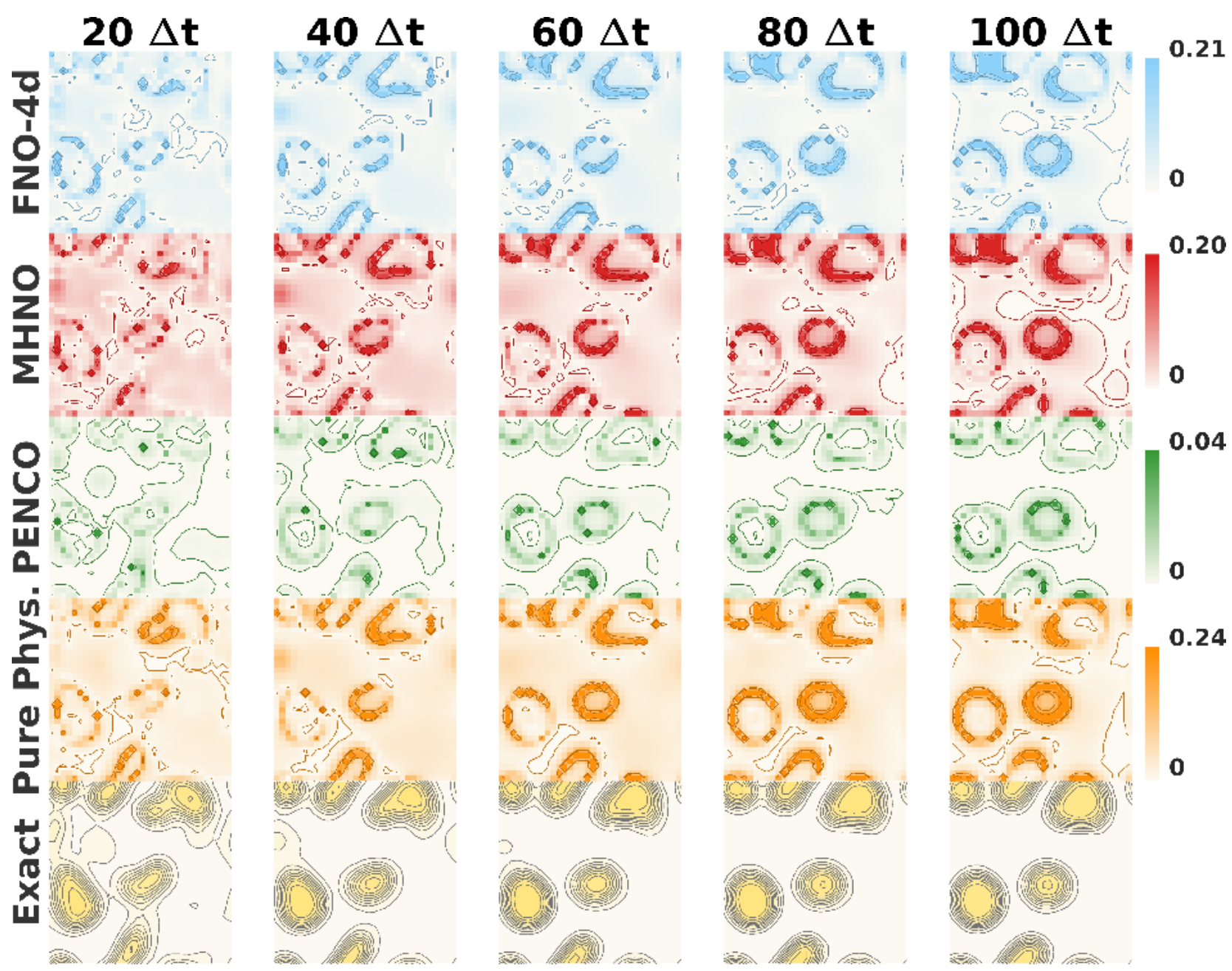

Figure 4-4: 2D pointwise relative error for the CH equation across FNO-4d, MHNO, PENCO, and the pure-physics baseline ($L = 2$, grid $32^3$, $\Delta t = 0.005$, $\lambda = 0.25$, $\epsilon = 0.05$, $N = 200$).

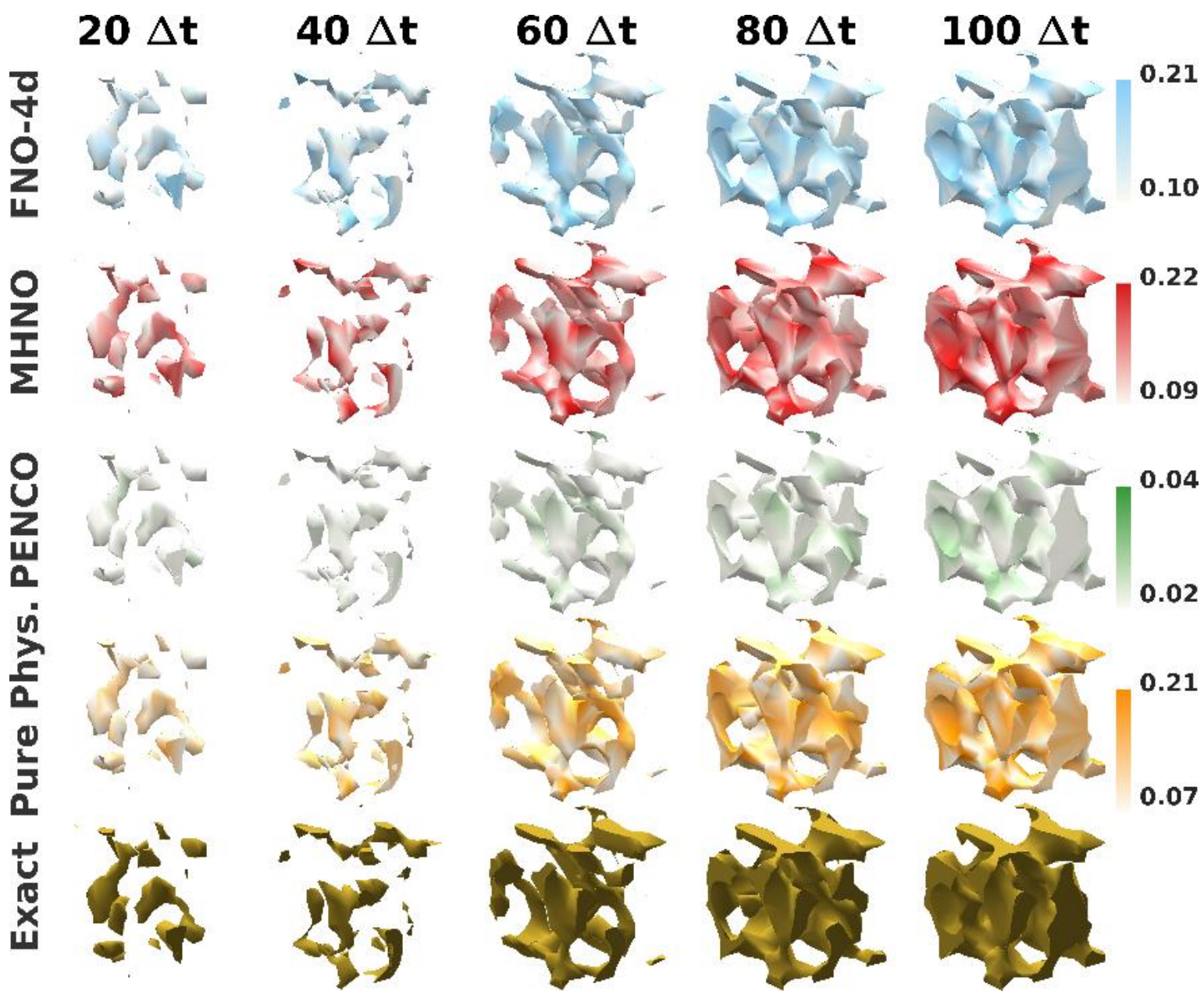

Figure 4-5: 3D iso-surface error visualization for the CH system (same setup as Figure 4.4).

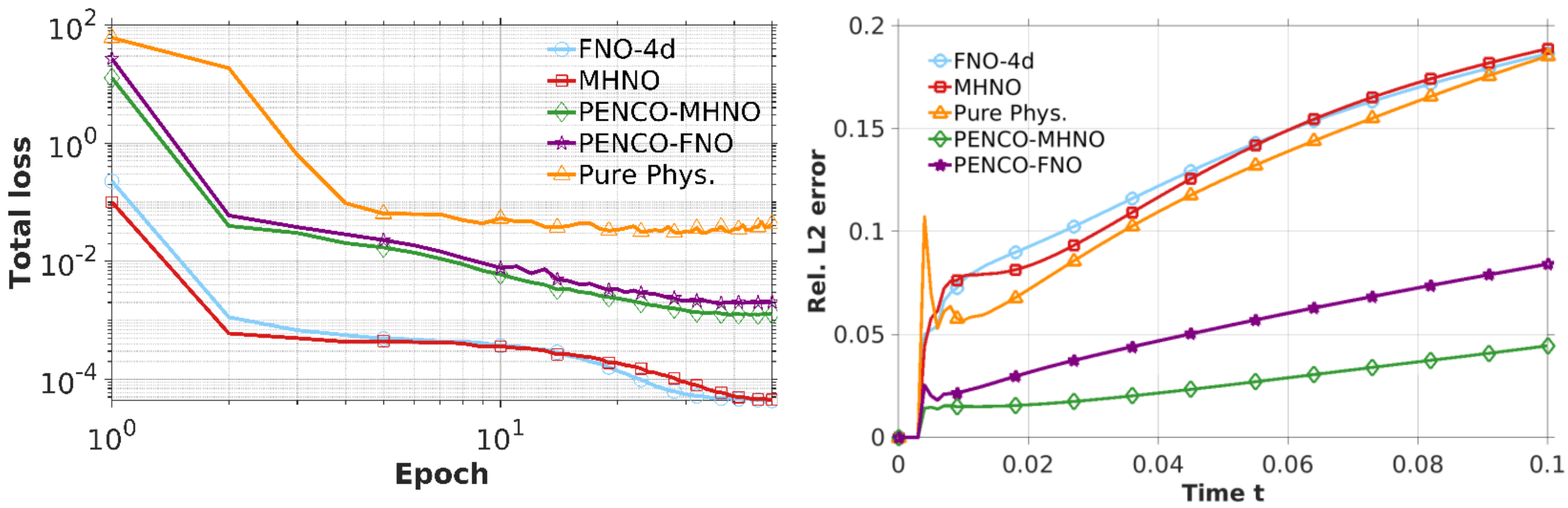

Figure 4-6: Total training loss and relative $L_2$-error for the CH system, showing the improved long-horizon accuracy of PENCO compared with all baseline models.

The numerical comparisons in Table 4-2 highlight how strongly the conservation constraint influences predictive accuracy. When the training set is very small ($\mathrm{N} = 50$), the three baseline models behave similarly and accumulate error at a comparable rate. The absence of a corrective physical mechanism allows mass imbalance and numerical diffusion to grow, which limits the achievable accuracy even when the architecture is expressive. Introducing physics regularization changes this behaviour immediately. PENCO reduces the long-horizon error by roughly **30–40 percent** compared with the baselines, revealing a clear gain in data efficiency even under the most restrictive conditions.

Increasing the training set to $\mathrm{N} = 100$ further exposes the separation between the methods. While FNO-4d and MHNO show only modest benefit from additional data, PENCO achieves errors that are **about**

**60 percent smaller** across the rollout. The hybrid model is the only configuration that translates the extra data into meaningful long-horizon improvement, demonstrating that accurate CH dynamics require both data and structural guidance.

With $\mathrm{N} = 200$, the contrast becomes even more apparent. The baseline methods plateau at nearly the same error levels as in the $\mathrm{N} = 100$ case, indicating limited data scalability for this conservative system. In comparison, PENCO reduces the error by **a factor of four to five**, showing that the physics term enables the network to extract and generalize information from the data far more effectively. The pure-physics model remains stable but lacks the refinement gained from data, placing it between the baselines and PENCO.

Table 4-2: Relative $L_2$ norm errors for CH equation ($L = 2$, grid $32^3$, $\Delta t = 0.005$, $\epsilon = 0.05$)

| N | Method | $0\Delta t$ | $20\Delta t$ | $40\Delta t$ | $60\Delta t$ | $80\Delta t$ | $100\Delta t$ |
|---|---|---|---|---|---|---|---|
| 50 | FNO-4d | 0 | 0.11 | 0.16 | 0.19 | 0.21 | 0.23 |
| 50 | MHNO | 0 | 0.19 | 0.16 | 0.17 | 0.19 | 0.21 |
| 50 | PENCO ($\lambda = 0.25$) | 0 | 0.08 | 0.11 | 0.14 | 0.16 | 0.19 |
| 50 | PENCO ($\lambda = 0.50$) | 0 | 0.08 | 0.11 | 0.14 | 0.17 | 0.19 |
| 50 | PENCO ($\lambda = 0.75$) | 0 | 0.08 | 0.11 | 0.14 | 0.16 | 0.18 |
| 50 | Pure Phys. ($\lambda = 1.0$) | 0 | 0.07 | 0.11 | 0.14 | 0.16 | 0.19 |
| 100 | FNO-4d | 0 | 0.12 | 0.15 | 0.18 | 0.19 | 0.2 |
| 100 | MHNO | 0 | 0.1 | 0.13 | 0.16 | 0.18 | 0.2 |
| 100 | PENCO ($\lambda = 0.25$) | 0 | 0.03 | 0.04 | 0.06 | 0.08 | 0.09 |
| 100 | PENCO ($\lambda = 0.50$) | 0 | 0.03 | 0.05 | 0.06 | 0.08 | 0.09 |
| 100 | PENCO ($\lambda = 0.75$) | 0 | 0.03 | 0.05 | 0.06 | 0.08 | 0.09 |
| 100 | Pure Phys. ($\lambda = 1.0$) | 0 | 0.07 | 0.11 | 0.14 | 0.16 | 0.19 |
| 200 | FNO-4d | 0 | 0.09 | 0.12 | 0.15 | 0.17 | 0.19 |
| 200 | MHNO | 0 | 0.08 | 0.12 | 0.15 | 0.17 | 0.19 |
| **200** | **PENCO ($\lambda = 0.25$)** | **0** | **0.02** | **0.02** | **0.03** | **0.04** | **0.04** |
| **200** | **PENCO ($\lambda = 0.50$)** | **0** | **0.02** | **0.02** | **0.03** | **0.04** | **0.04** |
| 200 | PENCO ($\lambda = 0.75$) | 0 | 0.02 | 0.02 | 0.03 | 0.04 | 0.05 |
| 200 | Pure Phys. ($\lambda = 1.0$) | 0 | 0.07 | 0.11 | 0.14 | 0.16 | 0.19 |

### 4.1.3 Spontaneous Pattern Formation: The Swift-Hohenberg Equation

The SH equation governs spontaneous pattern formation in systems where periodic structures emerge from initially uniform states. It plays a central role in the formation of convection rolls in heated fluids [63] and in characterizing wrinkle formation in thin elastic films [64]. In the unified framework this model is obtained by activating the $\alpha_3$ term, producing a dynamics governed by the interaction between a growth-inducing Laplacian and a stabilizing biharmonic operator. Their competition selects a preferred spatial frequency, leading to the striped or cellular morphologies characteristic of SH evolution.

The SH dynamics provide a stringent test for surrogate models because the preferred wavelength and phase of the stripes must remain coherent over long times. The error visualizations in Figure 4-7 and Figure 4-8 report the spatial distribution of the pointwise relative error between the predicted and reference solutions, highlighting the sensitivity of the system to small phase or curvature deviations. The data-driven baselines gradually exhibit increasing errors as the rollout proceeds, with deviations concentrating near defect regions and along misaligned stripe interfaces. This behavior is most pronounced for FNO-4D, which departs from the reference pattern earliest and with the largest phase shift. MHNO retains coherence longer but ultimately shows similar drift, since neither method explicitly enforces the wavelength-selection mechanism governing SH morphology.

The hybrid PENCO model ($\lambda = 0.25$) maintains significantly lower error levels throughout the evolution. By combining PDE-based stabilization with data-driven correction, it keeps the emergent wavelength locked while accurately resolving nonlinear interactions that pure physics alone cannot capture. The PENCO variant built on MHNO achieves the most consistent agreement with the reference across all time frames and preserves the fine-scale oscillatory structure of the pattern.

The purely physics-based configuration behaves differently in this system compared with others. After the first second of evolution its accuracy drops below that of the MHNO baseline. This decline is explained by the sensitivity of the SH equation to small phase and amplitude discrepancies. The PDE alone maintains global structure but cannot correct the fine nonlinear adjustments required to keep the stripes perfectly aligned. Without data supervision, small errors in curvature and phase accumulate until the predicted pattern slowly shifts away from the true solution.

Figure 4-9 shows that both data-driven models and the pure physics solver exhibit an initial error overshoot. The data-driven methods lack the wavelength selection mechanism of the SH equation, while the pure physics model lacks data supervision during the rapid growth phase. After this transient, FNO 4d and MHNO accumulate phase drift and their errors steadily increase, whereas the PENCO variants remain stable with consistently low errors. The pure physics model, although less accurate than MHNO for most of the rollout, gradually closes the gap and approaches the MHNO accuracy by the final time frame.

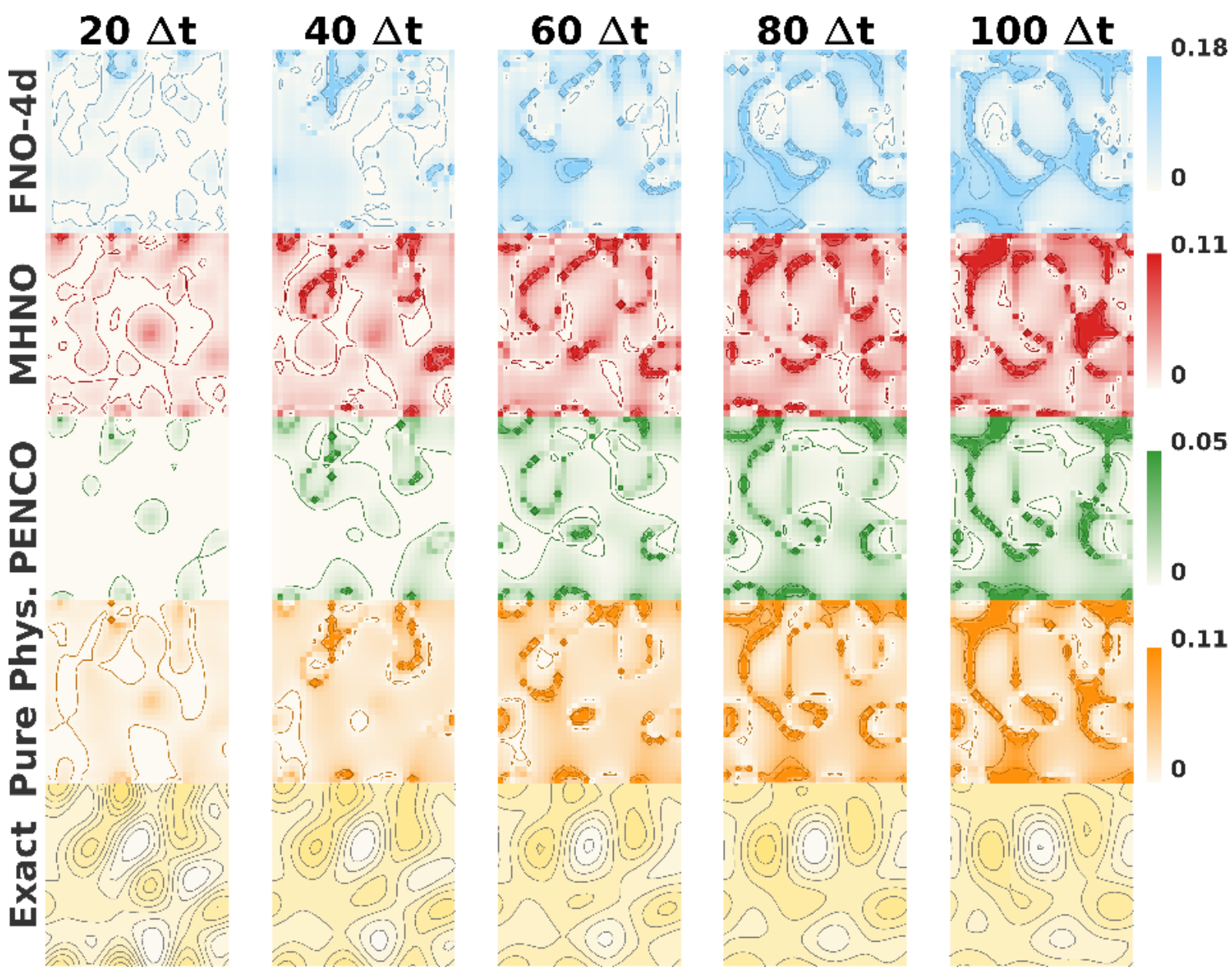

Figure 4-7: 2D pointwise relative error contours for the SH equation across different ($L = 15$, grid $32^3$, $\Delta t = 0.05$, $\epsilon = 0.15$, $N = 200$).

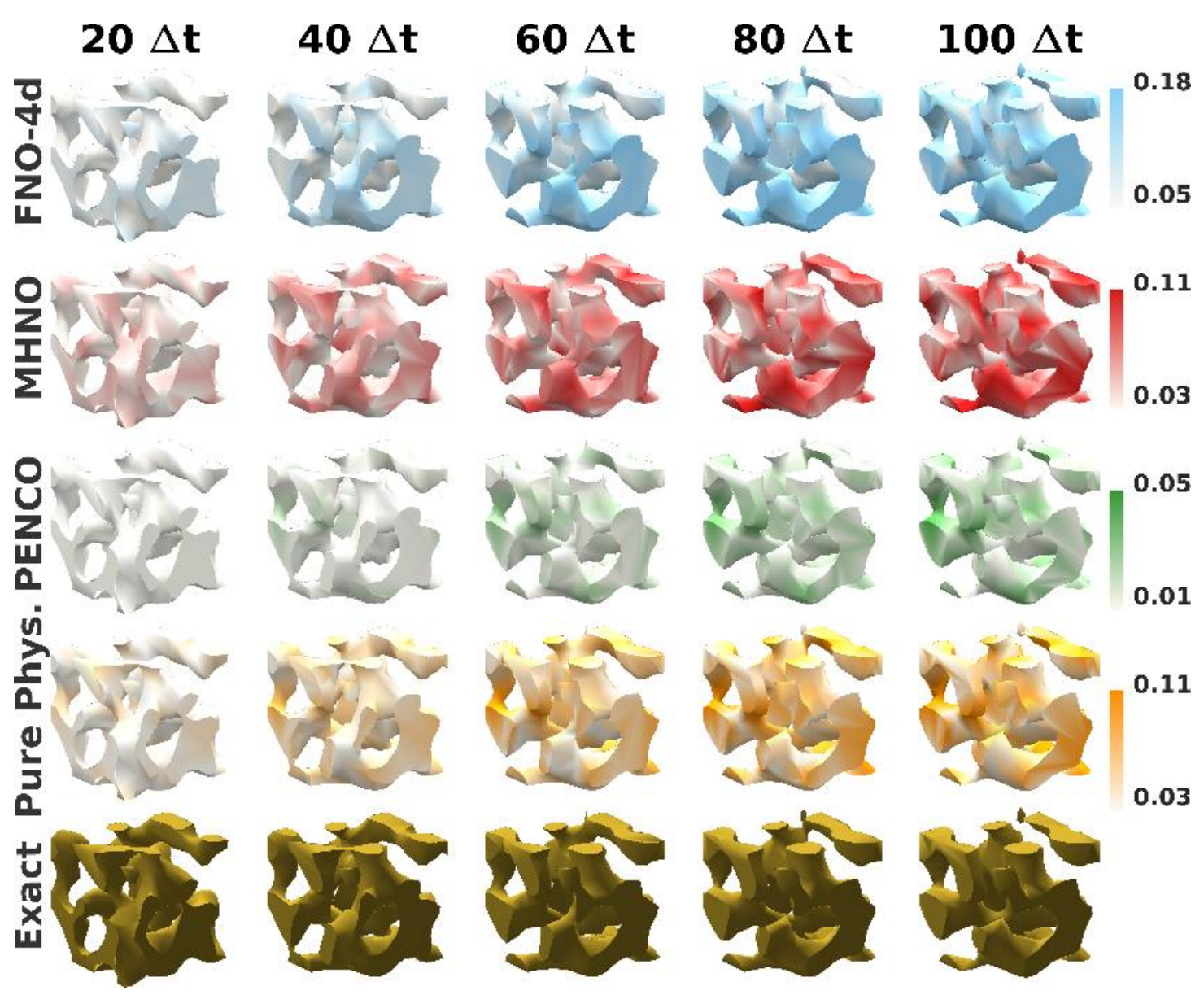

Figure 4-8: 3D iso-surface error rendering for the SH system, highlighting spatial deviations from the reference pattern (same setup as Figure 4.7)

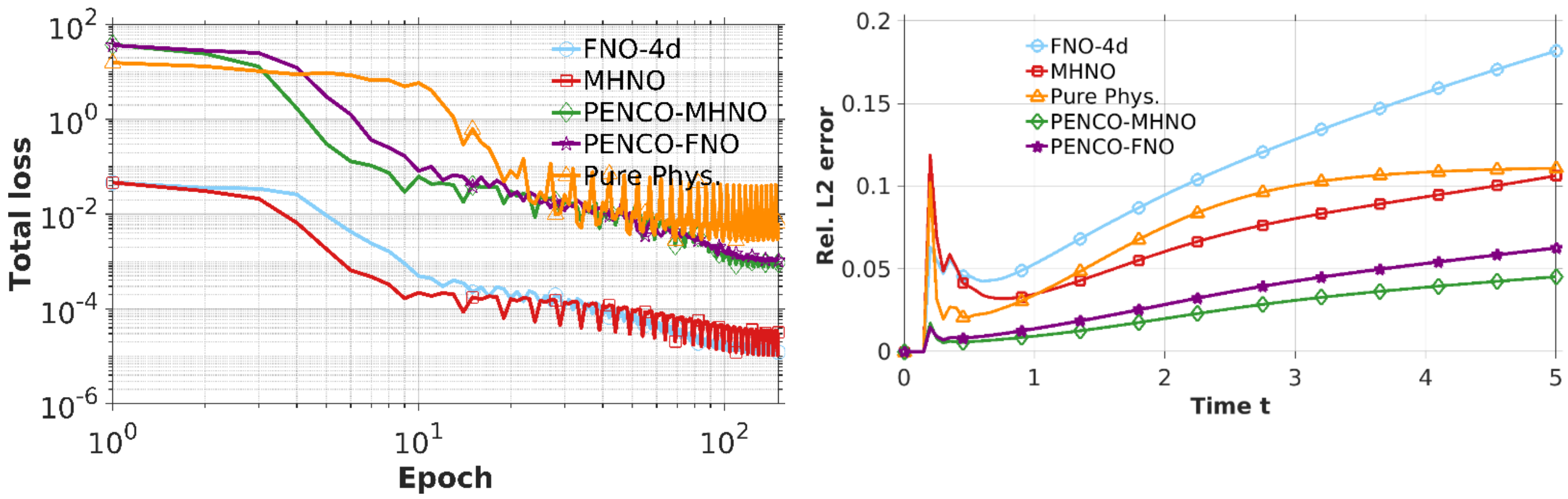

Figure 4-9: Total loss and relative $L_2$-error for the SH system showing stable long-horizon accuracy of the PENCO models and overshoot followed by drift in the data-driven baselines.

The quantitative results in Table 4-3 show that adding physics information produces a clear and systematic improvement across all dataset sizes. For $\mathrm{N} = 50$, the data-driven baselines settle at relatively large final errors, whereas the PENCO variants reduce the discrepancy by a factor of about two to three. This corresponds to an accuracy gain from roughly **60-70%**. The pure physics configuration also remains more reliable than both data-driven models, which demonstrates that structural constraints alone already strengthen stability in this pattern-forming regime.

At $\mathrm{N} = 100$, the advantage becomes even more visible. The data-driven networks experience only mild progress when more samples are provided, while PENCO continues to lower the error substantially. At the final time evaluation, the hybrid models outperform FNO-4D and MHNO by approximately **50 to**

**60 percent**. This behaviour indicates that the additional data are useful only when combined with physical guidance that keeps the dynamics aligned with the correct wavelength selection.

With N=200, all methods show improvement, yet the differences between learning strategies remain clear. The PENCO variants reach the lowest errors at every evaluation time and reduce the long-horizon discrepancy by a factor of **about three to four** when compared with FNO 4D. Relative to MHNO, the improvement is from **40-50 percent**. This level of stability across the entire rollout shows that PENCO delivers higher accuracy and better data efficiency, enabling it to reproduce the SH pattern formation with much stronger fidelity despite the sensitivity of this higher order system.

Table 4-3: Relative $L_2$ -norm errors for SH equation ($L = 15$, grid $32^3$, $\Delta t = 0.05$, $\epsilon = 0.15$)

| **N** | **Method** | $\mathbf{0\Delta t}$ | $\mathbf{20\Delta t}$ | $\mathbf{40\Delta t}$ | $\mathbf{60\Delta t}$ | $\mathbf{80\Delta t}$ | $\mathbf{100\Delta t}$ |
|---|---|---|---|---|---|---|---|
| 50 | FNO-4d | 0 | 0.12 | 0.17 | 0.21 | 0.23 | 0.24 |
| 50 | MHNO | 0 | 0.11 | 0.16 | 0.18 | 0.21 | 0.25 |
| 50 | PENCO ($\lambda = 0.25$) | 0 | 0.03 | 0.05 | 0.06 | 0.07 | 0.08 |
| 50 | PENCO ($\lambda = 0.50$) | 0 | 0.03 | 0.06 | 0.07 | 0.08 | 0.08 |
| 50 | PENCO ($\lambda = 0.75$) | 0 | 0.03 | 0.06 | 0.08 | 0.09 | 0.09 |
| 50 | Pure Phys. ($\lambda = 1.0$) | 0 | 0.03 | 0.08 | 0.1 | 0.11 | 0.11 |
| 100 | FNO-4d | 0 | 0.07 | 0.11 | 0.14 | 0.15 | 0.15 |
| 100 | MHNO | 0 | 0.07 | 0.1 | 0.12 | 0.14 | 0.17 |
| 100 | PENCO ($\lambda = 0.25$) | 0 | 0.02 | 0.04 | 0.05 | 0.06 | 0.07 |
| 100 | PENCO ($\lambda = 0.50$) | 0 | 0.02 | 0.04 | 0.06 | 0.07 | 0.07 |
| 100 | PENCO ($\lambda = 0.75$) | 0 | 0.02 | 0.05 | 0.07 | 0.08 | 0.09 |
| 100 | Pure Phys. ($\lambda = 1.0$) | 0 | 0.03 | 0.08 | 0.1 | 0.11 | 0.11 |
| 200 | FNO-4d | 0 | 0.05 | 0.09 | 0.13 | 0.16 | 0.18 |
| 200 | MHNO | 0 | 0.03 | 0.06 | 0.08 | 0.09 | 0.11 |
| **200** | **PENCO ($\boldsymbol{\lambda = 0.25}$)** | **0** | **0.01** | **0.02** | **0.03** | **0.04** | **0.05** |
| 200 | PENCO ($\lambda = 0.50$) | 0 | 0.01 | 0.02 | 0.04 | 0.05 | 0.06 |
| 200 | PENCO ($\lambda = 0.75$) | 0 | 0.01 | 0.03 | 0.05 | 0.06 | 0.07 |
| 200 | Pure Phys. ($\lambda = 1.0$) | 0 | 0.03 | 0.08 | 0.1 | 0.11 | 0.11 |

### 4.1.4 Atomic-Scale Crystallization: The Phase Field Crystal Model

The Phase Field Crystal equation models crystallization at atomic resolution and provides a continuum description of periodic density fields. It captures essential physical mechanisms such as lattice formation, defect migration and long-range diffusion that governs microstructural stability. The equation combines a SH-type pattern-forming energy with a mass-conserving CH dynamics, giving rise to periodic structures that evolve through conserved fluxes [65, 66]. Within our unified formulation this behavior is obtained by activating the $\alpha_4$ term, which introduces the wavelength-selective energy needed to generate a crystalline lattice.

The spatial error distributions for the PFC system are reported in Figure 4-10 and Figure 4-11, where the pointwise relative error between predicted and reference solutions is shown in two and three dimensions, respectively. All models reproduce the interior lattice structure reasonably well, while the clearest differences emerge near the domain boundaries, where conserved flux and long-range coupling play a dominant role. The data-driven networks gradually exhibit increasing deviations in these regions, consistent with the relative error trends in Figure 4-12. Both FNO-4d and MHNO show an early rise in error followed by steady growth, indicating that they do not fully capture the conserved transport and wavelength selection characteristic of the PFC dynamics.

The hybrid PENCO model maintains stable lattice alignment across the entire domain and avoids the boundary drift observed in the data-driven approaches. The error curve in Figure 4-12 reflects this stability, remaining low and slowly varying throughout the rollout. The purely physics-based solution retains global consistency but lacks the refinement provided by training data. The overall behavior resembles the trends seen in the AC case, which can be attributed to the shared structure of the conserved flux operator and the similar number of update steps used during training. These common elements lead to comparable stability characteristics across both systems, while the hybrid PENCO design continues to deliver the most accurate representation of the evolving patterns.

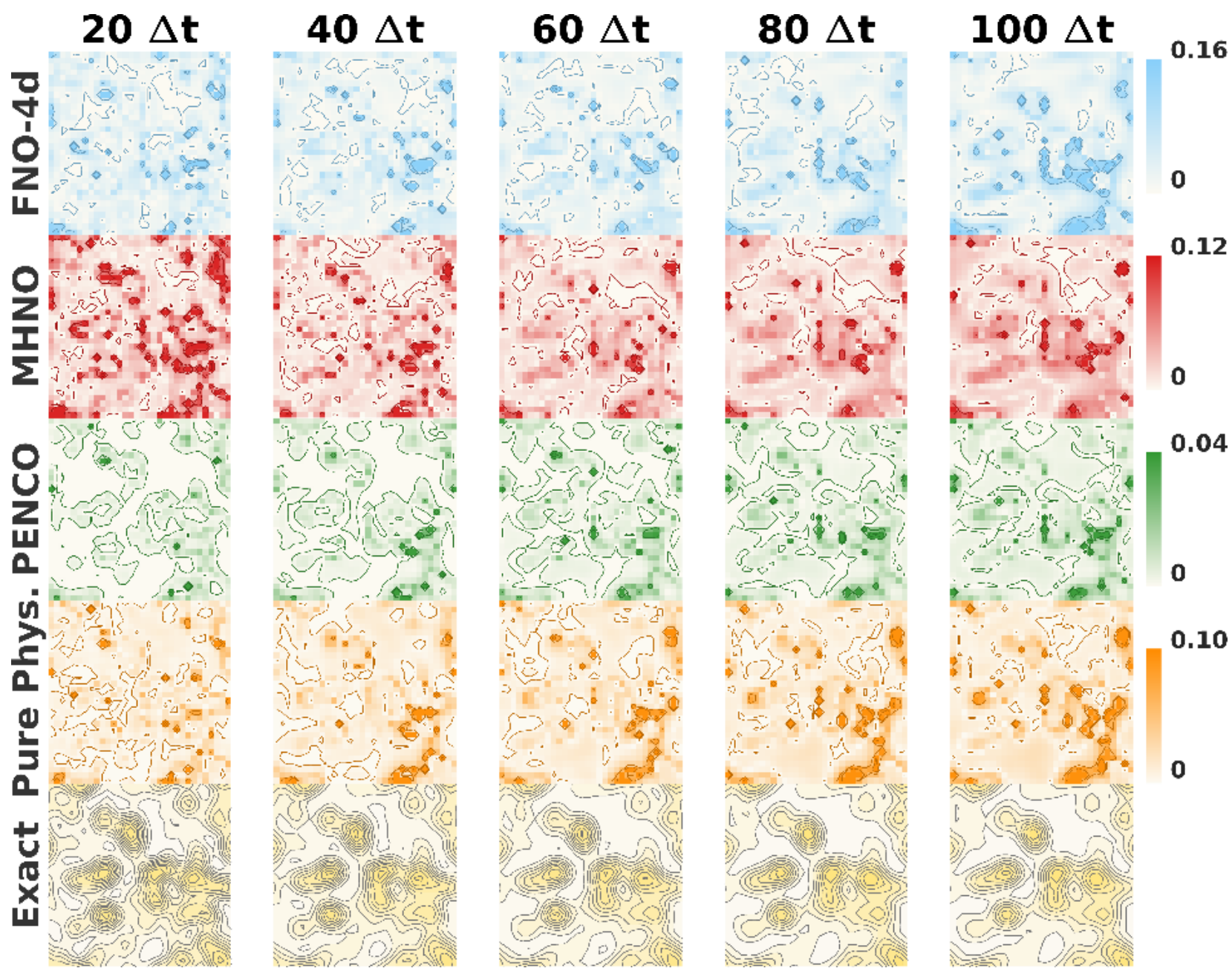

Figure 4-10: Two-dimensional contour evolution for the PFC system ($L = 10\pi$, grid $32^3$, $\Delta t = 0.01$, $\epsilon = 0.5$, $\mathrm{N} = 200$)

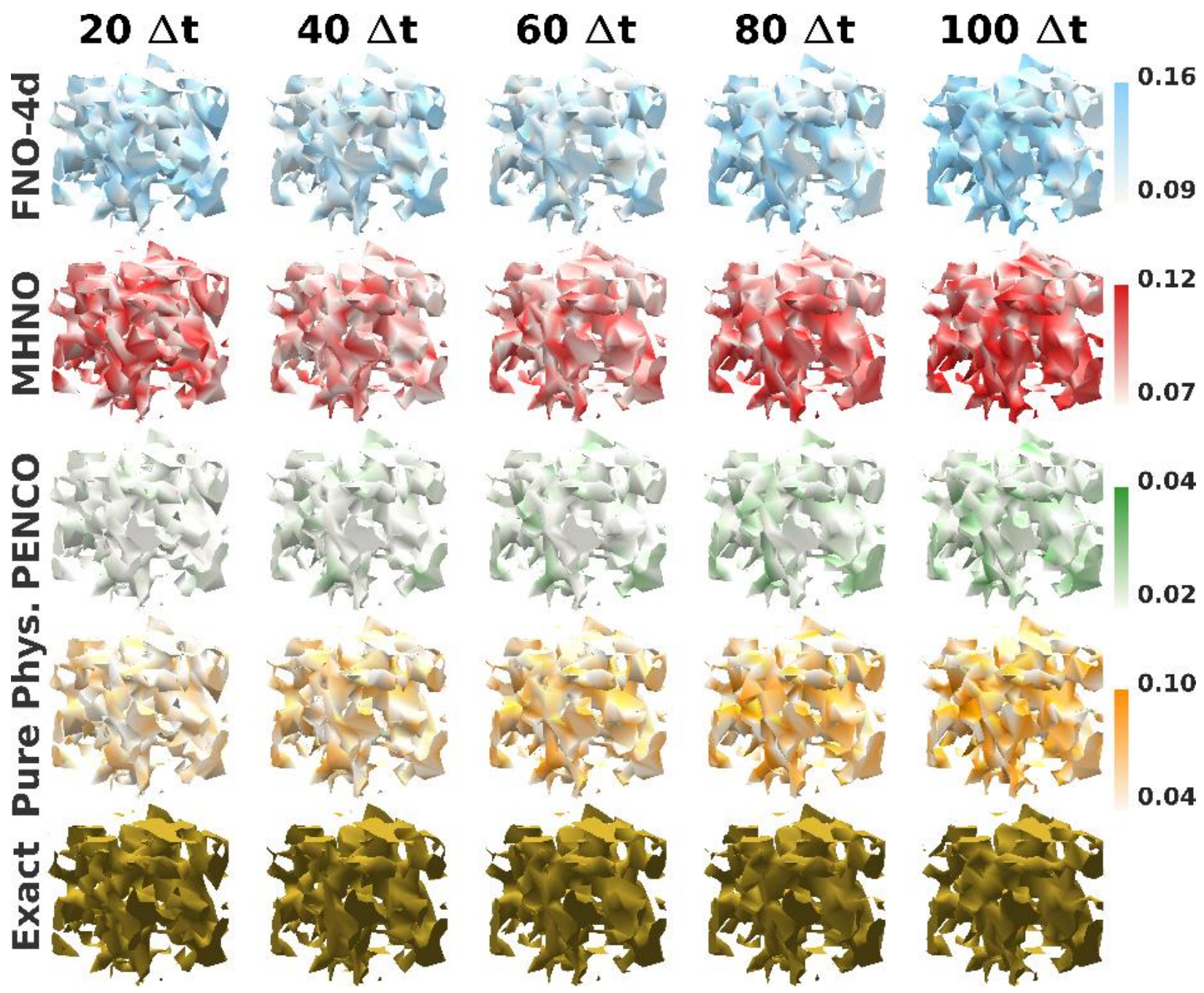

Figure 4-11: Three-dimensional crystalline morphology evolution across all models.

Figure 4-12 compares the total training loss and the temporal evolution of the relative $L_2$-error for all methods. The results clearly demonstrate the advantage of coupling data with physics. The hybrid PENCO configurations display smooth and steady optimization behavior, while purely data-driven and purely physics-based models show larger oscillations and slower error decay.

The relative error plots highlight a distinct contrast between learning regimes. After the initial transient, the FNO-4D and MHNO models undergo a sharp error rise, followed by a gradual stabilization at significantly higher magnitudes. This indicates unstable transient dynamics caused by the absence of explicit conservation and energy-dissipation constraints. In contrast, the PENCO variants maintain consistently low error levels across the full rollout. The inclusion of even moderate physics weighting effectively suppresses the early overshoot and prevents the runaway divergence characteristic of the purely data-based networks. The fully physics-based model also overshoots initially due to the lack of data, and its error increases gently over time, though it still outperforms purely data-driven models. PENCO achieves the lowest and most stable errors overall, highlighting its superior handling of complex conservative dynamics.

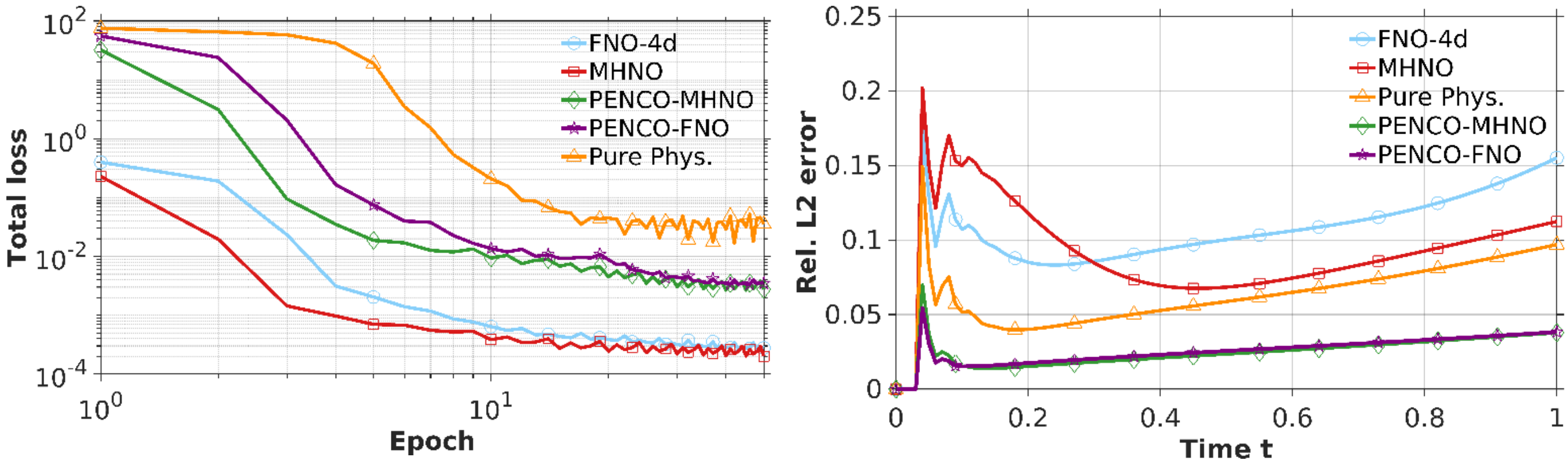

Figure 4-12: Total loss and relative error for the PFC system showing the accuracy and stability of the different models.

The results in Table 4-4 show that PENCO achieves a clear accuracy advantage at every data level. For $\mathrm{N} = 50$, the hybrid models reduce the long-horizon error by about **70–75%** compared with both data-driven baselines, demonstrating that even modest physics guidance substantially improves stability when training data are scarce.

At $\mathrm{N} = 100$, PENCO maintains this lead, lowering the final error by roughly **65–75%** relative to the baselines. The pure-physics model also surpasses the data-driven methods, though it remains less accurate than the hybrid configurations, indicating the added benefit of data-based refinement.

For $\mathrm{N} = 200$, all models improve with additional data, yet PENCO continues to deliver the smallest errors, achieving a **50–75%** reduction compared with the data-driven architectures. These gains indicate that the hybrid formulation retains the essential physical structure of the PFC system more effectively than either data-only or physics-only learning.

Table 4-4: Relative $L_2$-norm errors for the PFC equation ($L = 10\pi$, grid $32^3$, $\Delta t = 0.01$, $\epsilon = 0.5$)

| N | Method | $0\Delta t$ | $20\Delta t$ | $40\Delta t$ | $60\Delta t$ | $80\Delta t$ | $100\Delta t$ |
|---|---|---|---|---|---|---|---|
| 50 | FNO-4d | 0 | 0.17 | 0.21 | 0.25 | 0.3 | 0.35 |
| 50 | MHNO | 0 | 0.14 | 0.18 | 0.24 | 0.31 | 0.38 |
| 50 | PENCO ($\lambda = 0.25$) | 0 | 0.04 | 0.05 | 0.07 | 0.08 | 0.1 |
| 50 | PENCO ($\lambda = 0.50$) | 0 | 0.04 | 0.05 | 0.07 | 0.08 | 0.1 |
| 50 | PENCO ($\lambda = 0.75$) | 0 | 0.04 | 0.05 | 0.07 | 0.08 | 0.1 |
| 50 | Pure Phys. ($\lambda = 1.0$) | 0 | 0.04 | 0.05 | 0.06 | 0.08 | 0.1 |
| 100 | FNO-4d | 0 | 0.09 | 0.11 | 0.14 | 0.18 | 0.24 |
| 100 | MHNO | 0 | 0.12 | 0.09 | 0.1 | 0.12 | 0.15 |
| 100 | PENCO ($\lambda = 0.25$) | 0 | 0.02 | 0.03 | 0.04 | 0.05 | 0.06 |
| 100 | PENCO ($\lambda = 0.50$) | 0 | 0.02 | 0.03 | 0.04 | 0.05 | 0.06 |
| 100 | PENCO ($\lambda = 0.75$) | 0 | 0.02 | 0.03 | 0.04 | 0.05 | 0.06 |
| 100 | Pure Phys. ($\lambda = 1.0$) | 0 | 0.04 | 0.05 | 0.06 | 0.08 | 0.1 |
| 200 | FNO-4d | 0 | 0.09 | 0.09 | 0.11 | 0.12 | 0.16 |
| 200 | MHNO | 0 | 0.12 | 0.07 | 0.07 | 0.09 | 0.11 |
| **200** | **PENCO ($\boldsymbol{\lambda = 0.25}$)** | **0** | **0.02** | **0.02** | **0.03** | **0.03** | **0.04** |
| **200** | **PENCO ($\boldsymbol{\lambda = 0.50}$)** | **0** | **0.01** | **0.02** | **0.03** | **0.03** | **0.04** |
| 200 | PENCO ($\lambda = 0.75$) | 0 | 0.02 | 0.02 | 0.03 | 0.03 | 0.04 |
| 200 | Pure Phys. ($\lambda = 1.0$) | 0 | 0.04 | 0.05 | 0.06 | 0.08 | 0.1 |

### 4.1.5 Thin-Film Epitaxial Growth: The Molecular Beam Epitaxy Model

The Molecular Beam Epitaxy equation models thin-film growth driven by the interplay of high-order surface diffusion and a nonlinear slope-selection flux, a mechanism central to understanding mound formation and coarsening in epitaxial systems [63, 64]. Within our unified formulation this behavior is obtained by activating the $\alpha_5$ term, which introduces the characteristic fourth order smoothing and slope-dependent dynamics that make the MBE system one of the most challenging benchmarks in our study.

The MBE case reveals the clearest separation between approaches. As shown by the two- and three-dimensional error visualizations in Figure 4-13 and Figure 4-14, the data-driven models rapidly exhibit large deviations from the reference solution, with errors concentrating near mound interfaces and regions of strong slope variation. In contrast, the pure-physics model remains more stable but still lacks fine-scale accuracy. These qualitative differences are confirmed quantitatively by Figure 4-15, where FNO-4d and MHNO show a sharp early error surge and continued growth, indicating that they cannot sustain the nonlinear slope-selection mechanism of the MBE equation. In contrast, the hybrid PENCO model remains closely aligned with the reference across the entire rollout, accurately preserving both the surface morphology and boundary motion. Its relative error stays low and nearly flat, reflecting a level of stability that is difficult to achieve for an equation governed by high-order diffusion and strongly nonlinear slope-selection fluxes.

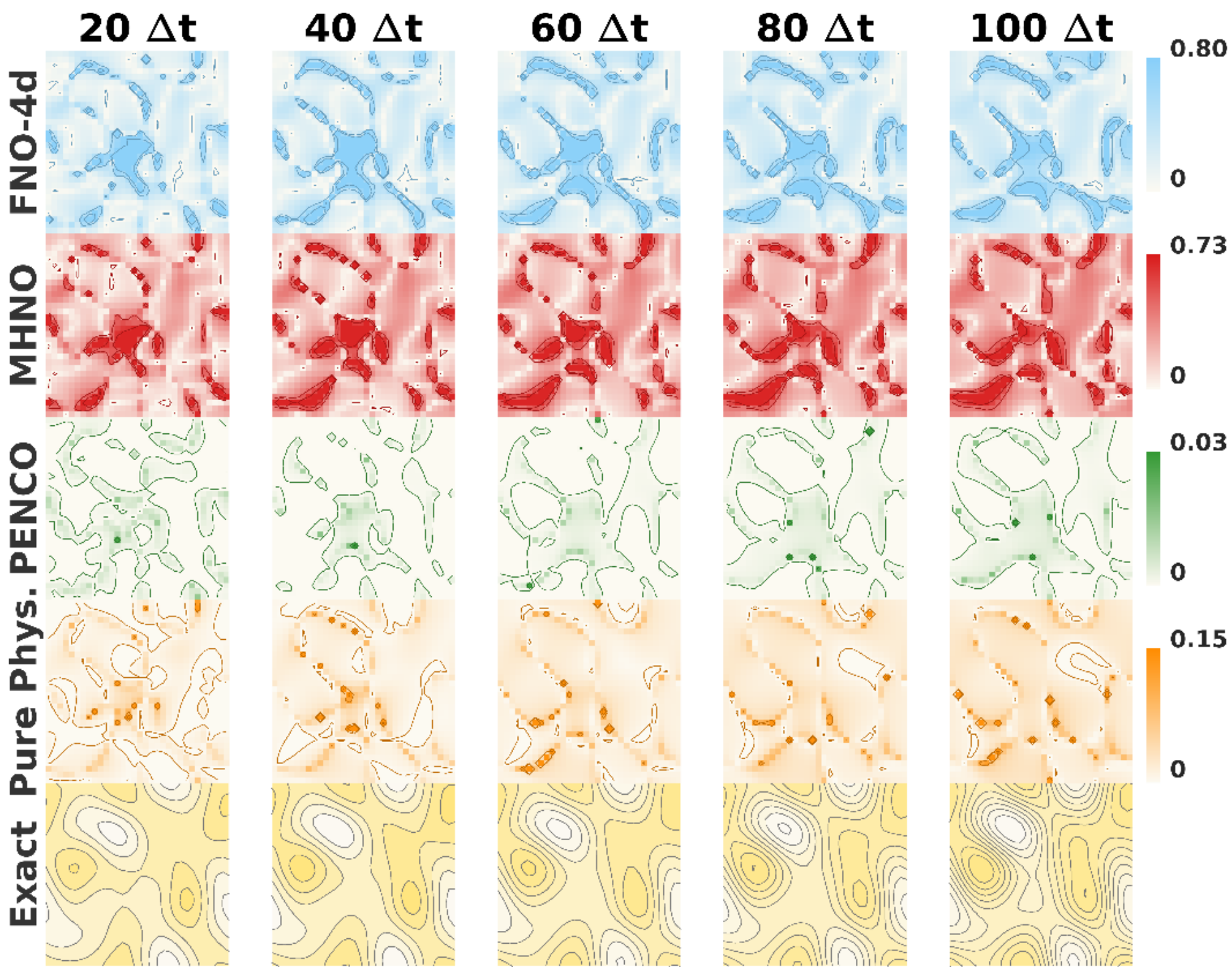

Figure 4-13: Two-dimensional contour evolution of the MBE surface field for FNO-4D, MHNO, PENCO, pure-physics, and the reference solution ($32^3$ *grid,* $\Delta t \ = \ 0.005$*,* $\epsilon = 0.1$*,* $N_t \ = 100$ *,* $L = 2\pi$*,* $N \ = 200$*).*

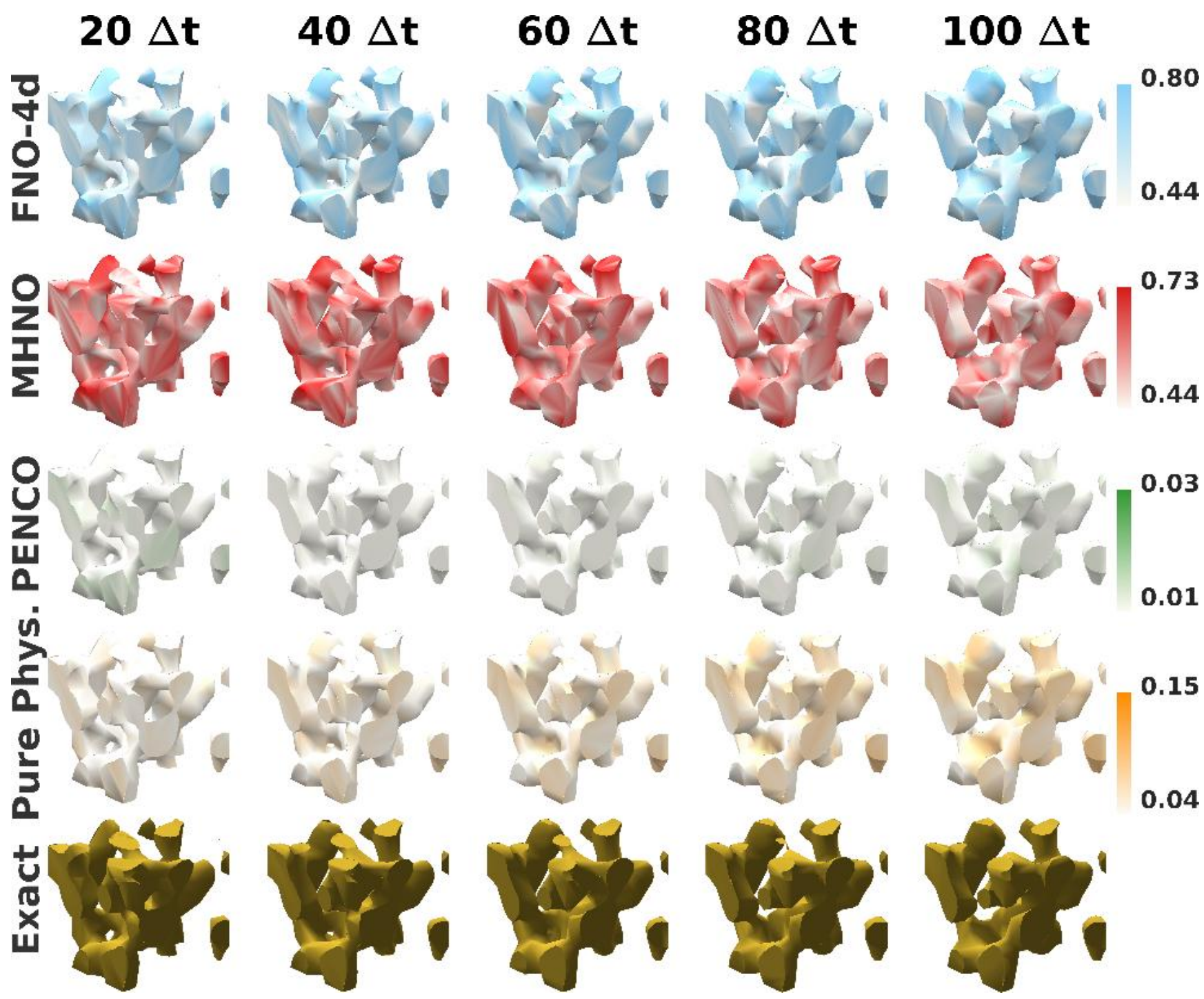

Figure 4-14: 3D phase evolution of the MBE system predicted by different models.

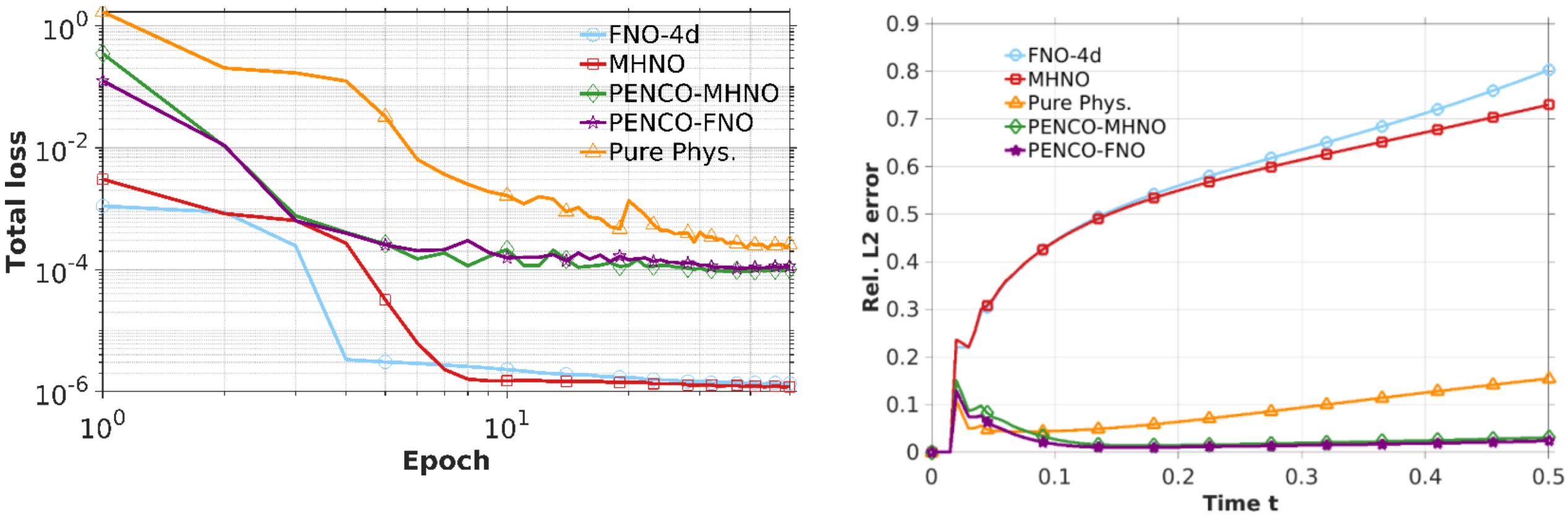

Figure 4-15: Total loss and relative $L_2$-norm error for the MBE equation using different methods

The quantitative results show the steep difficulty of the MBE system for data-driven surrogates. As summarized in Table 4-5, the purely data-driven models experience severe long-horizon degradation, while PENCO maintains substantially higher fidelity across all dataset sizes. With $\mathrm{N} = 50$, the discrepancy between the approaches is dramatic. PENCO reduces the long-term error by more than **95 %** relative to FNO-4D and by approximately **70 %** relative to MHNO, demonstrating its ability to stabilize the evolution even under minimal data availability. The pure-physics model also improves significantly over the data-driven baselines but remains about **30 %** less accurate than PENCO, reflecting the importance of data-guided correction for this strongly nonlinear, high-order PDE.

For $\mathrm{N} = 100$, the same ordering persists. PENCO lowers the final-time error by around 85 % compared with MHNO and by over **90 %** compared with FNO-4D, indicating that additional data amplify, rather than diminish, the benefit of physics-guided learning. With $\mathrm{N} = 200$, all models improve due to increased training samples, yet PENCO still provides the clearest advantage, achieving an accuracy gain of nearly **90 %** relative to MHNO and approximately **95 %** relative to FNO-4D. The pure-physics solution remains stable throughout but continues to trail PENCO by roughly **25–30 %**, highlighting the essential role of hybrid learning in capturing both the nonlinear flux term and the high-order diffusion that govern MBE surface evolution.

Table 4-5: Relative $L_2$-norm errors for MBE equation ($L = 2\pi$, grid $32^3$, $\Delta t = 0.005$, $\epsilon = 0.1$)

| N | Method | $0\Delta t$ | $20\Delta t$ | $40\Delta t$ | $60\Delta t$ | $80\Delta t$ | $100\Delta t$ |
|---|---|---|---|---|---|---|---|
| 50 | FNO-4d | 0 | 0.49 | 0.8 | 1.75 | 4.32 | 8.23 |
| 50 | MHNO | 0 | 0.33 | 0.45 | 0.54 | 0.61 | 0.66 |
| 50 | PENCO ($\lambda = 0.25$) | 0 | 0.08 | 0.09 | 0.13 | 0.17 | 0.21 |
| 50 | PENCO ($\lambda = 0.50$) | 0 | 0.06 | 0.08 | 0.11 | 0.15 | 0.18 |
| 50 | PENCO ($\lambda = 0.75$) | 0 | 0.05 | 0.07 | 0.1 | 0.13 | 0.17 |
| 50 | Pure Phys. ($\lambda = 1.0$) | 0 | 0.04 | 0.06 | 0.09 | 0.12 | 0.15 |
| 100 | FNO-4d | 0 | 0.46 | 0.62 | 0.8 | 1.16 | 1.74 |
| 100 | MHNO | 0 | 0.43 | 0.53 | 0.59 | 0.64 | 0.69 |
| 100 | PENCO ($\lambda = 0.25$) | 0 | 0.04 | 0.04 | 0.06 | 0.07 | 0.09 |
| 100 | PENCO ($\lambda = 0.50$) | 0 | 0.03 | 0.04 | 0.05 | 0.07 | 0.08 |
| 100 | PENCO ($\lambda = 0.75$) | 0 | 0.03 | 0.04 | 0.05 | 0.06 | 0.08 |
| 100 | Pure Phys. ($\lambda = 1.0$) | 0 | 0.04 | 0.06 | 0.09 | 0.12 | 0.15 |
| 200 | FNO-4d | 0 | 0.44 | 0.56 | 0.64 | 0.71 | 0.8 |
| 200 | MHNO | 0 | 0.44 | 0.55 | 0.61 | 0.67 | 0.73 |
| **200** | **PENCO ($\lambda = 0.25$)** | **0** | **0.03** | **0.01** | **0.02** | **0.02** | **0.03** |
| **200** | **PENCO ($\lambda = 0.50$)** | **0** | **0.02** | **0.02** | **0.02** | **0.03** | **0.03** |
| 200 | PENCO ($\lambda = 0.75$) | 0 | 0.02 | 0.02 | 0.02 | 0.03 | 0.04 |
| 200 | Pure Phys. ($\lambda = 1.0$) | 0 | 0.04 | 0.06 | 0.09 | 0.12 | 0.15 |

## 4.2 Out-of-Distribution (OOD) Generalization

To evaluate generalization beyond the GRF-based training distribution, we test all models on deterministic initial conditions whose geometry and spectral content differ fundamentally from the training set. For AC, CH, and SH we impose the smooth spherical initialization

$$u(\boldsymbol{x}, 0) = \tanh\left(\frac{R - \sqrt{x^2 + y^2 + z^2}}{\sqrt{b}\epsilon}\right), \tag{30}$$

while for the PFC system we introduce a strongly anisotropic star-shaped interface,

$$R_\theta = 10.0 + 5.0 \times \cos(6\theta), \quad u_0(\boldsymbol{x}) = \tanh[(R_\theta - d)/(7\epsilon)] \tag{31}$$

and for MBE we apply the torus configuration

$$u(x, 0) = \tanh\left(\frac{r_0 - \sqrt{\left(\sqrt{x^2 + y^2} - R\right)^2 + z^2}}{w}\right), \tag{32}$$

where $R$ is the major radius, $r_0$ is the tube radius, and $w$ denotes the interfacial width. These geometries, shown in Figure 4-16(a–e), vary in curvature, interface width, and topological complexity, making prediction accuracy highly sensitive to the underlying surface shape and therefore an informative probe of OOD generalization. All comparisons use $N = 200$ and the hybrid model with $\lambda = 0.25$.

Across all PDEs and OOD configurations, FNO-4D remains the least accurate method (as shown in Figure 4-16). Its MAE consistently grows throughout the rollout and tends to rise more rapidly than the other models once the evolution departs from the GRF statistics used during training. This reflects its limited ability to adapt when the unseen initial condition introduces geometric structure, sharper or smoother interfaces, or curvature patterns not represented in the training distribution.

MHNO consistently outperforms FNO-4D across all OOD tests. Its design includes a shared spectral encoder together with time-step–specific projection and transition networks, allowing it to refine spatial features dynamically and propagate information more coherently through time. These capabilities make MHNO more resilient to unseen geometries and non-GRF interfaces. In the AC and PFC systems, MHNO even surpasses the pure-physics model because these equations rely strongly on interface regularity and periodic microstructure, which benefit from MHNO's learned spatial refinement. In the remaining PDEs,

MHNO remains clearly stronger than FNO-4D, reflecting its improved ability to represent nonlinear evolution and sharper interface motion.

The hybrid PENCO models provide the strongest overall OOD generalization across four of the five systems, as observed in Figure 4-16. Physics regularization removes the early overshoot characteristic of the data-only operators and stabilizes long-horizon predictions when the initial condition departs significantly from the GRF statistics. By enforcing the structural constraints of each PDE while still benefiting from data-driven refinement, PENCO produces the most reliable MAE behavior for the spherical, star, and torus tests. In CH, the pure-physics model attains slightly higher accuracy because the dynamics are dominated by strict mass conservation and curvature-driven coarsening, both fully encoded in the PDE. In the remaining systems, particularly SH, PFC, and MBE, PENCO model achieves the best performance, as the combination of physics constraints and learned spatial refinement allows it to track wavelength-selective dynamics and nonlinear flux behavior more accurately than either data-only or physics-only models.

The pure-physics configuration behaves differently depending on the governing equation. It performs exceptionally well in CH, SH, and MBE, where the evolution is dominated by strict conservation laws, curvature-controlled coarsening, or slope-regulated diffusion; all of these mechanisms are fully encoded in the PDE and therefore generalize well to unseen initial conditions without needing data-driven correction. In AC and PFC, however, pure physics is matched or slightly overtaken by MHNO because these systems involve interface sharpening and periodic lattice formation that benefit from learned spatial detail not explicitly enforced by the PDE. Overall, the physics-only model remains stable in every OOD test but lacks the adaptive refinement provided by PENCO or MHNO when geometric complexity is high.

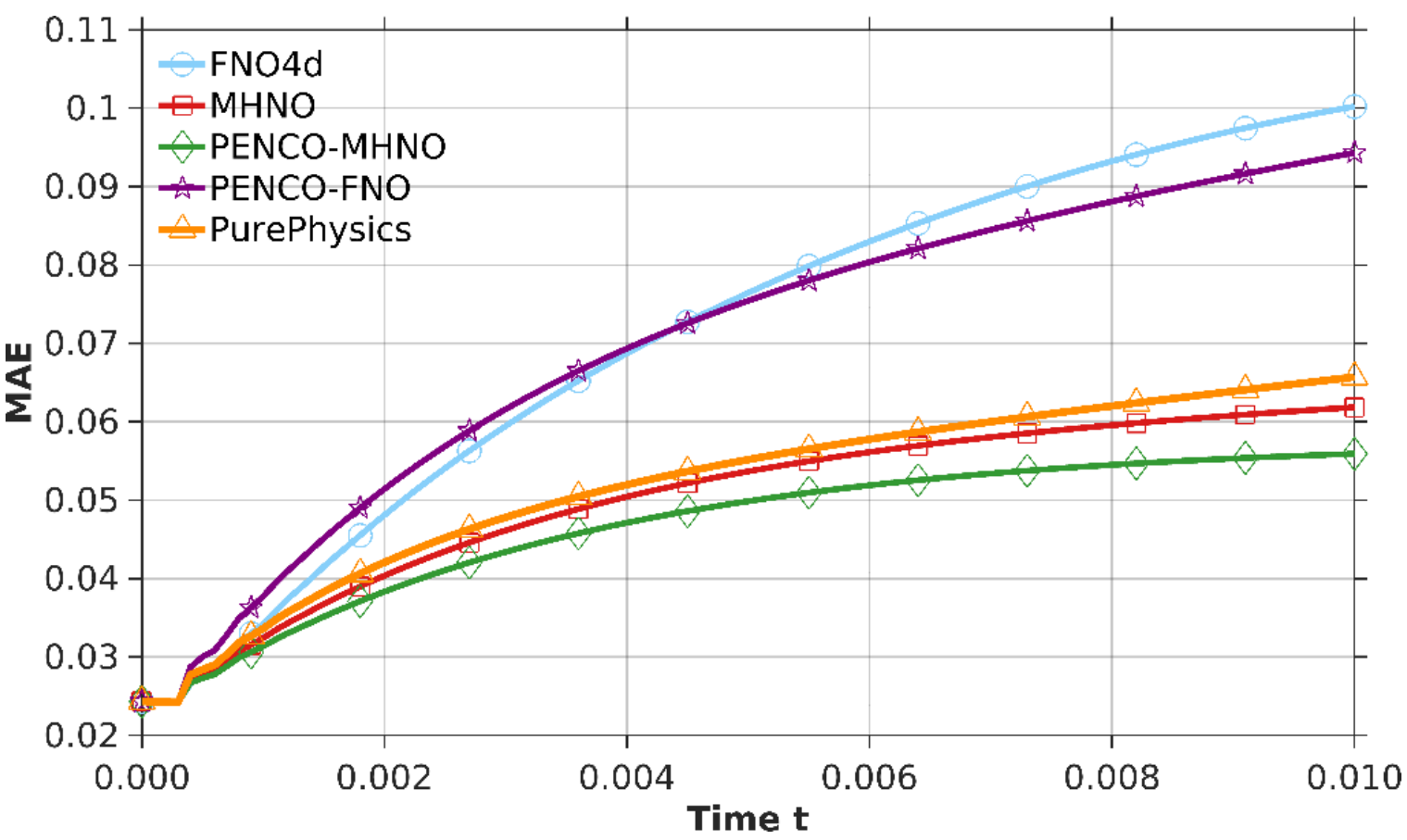

a) AC with spherical IC ($R = 0.5$, $\epsilon = 0.05$, $b = 2$)

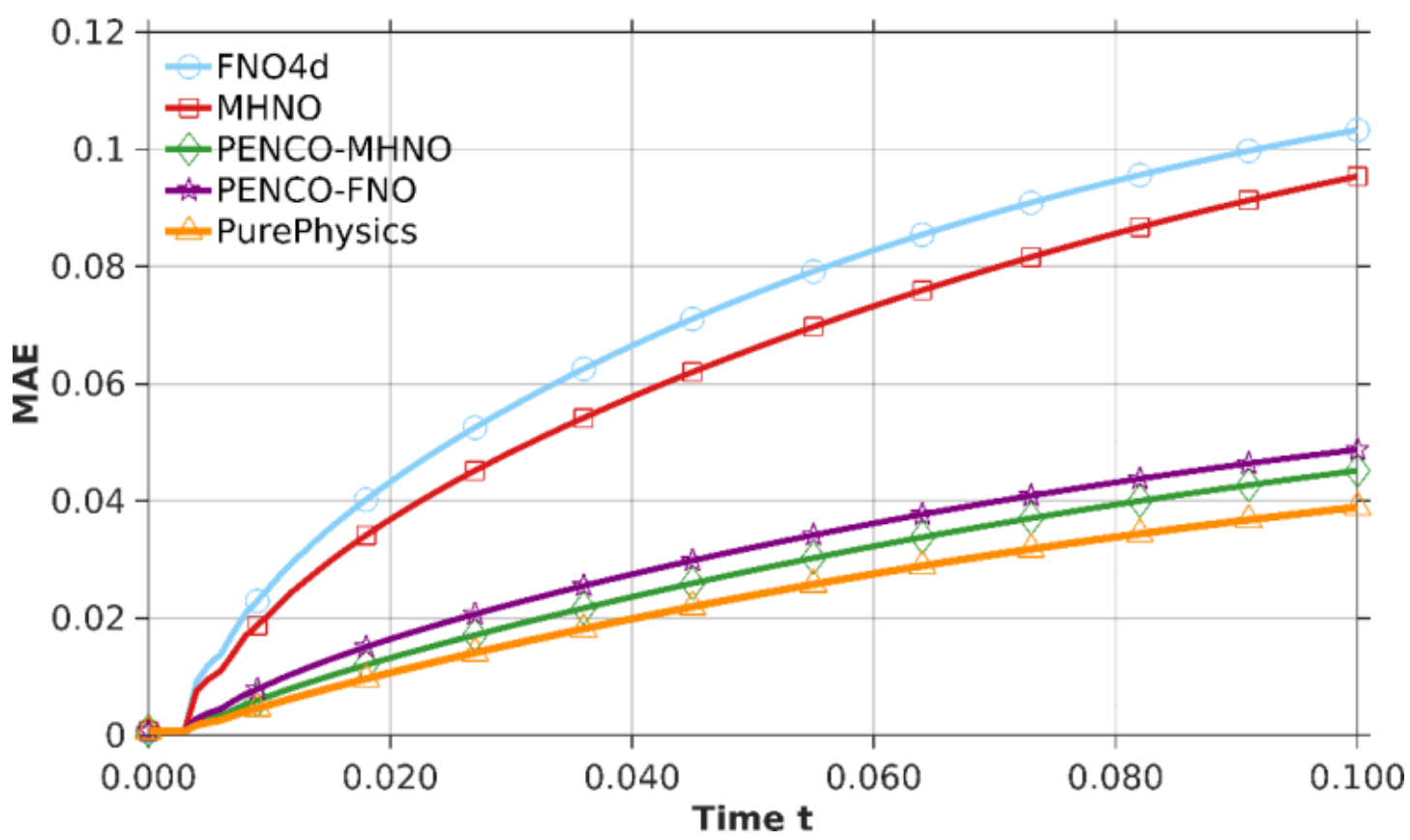

b) CH with spherical IC ($R = 0.5$, $\epsilon = 0.05$, $b = 2$)

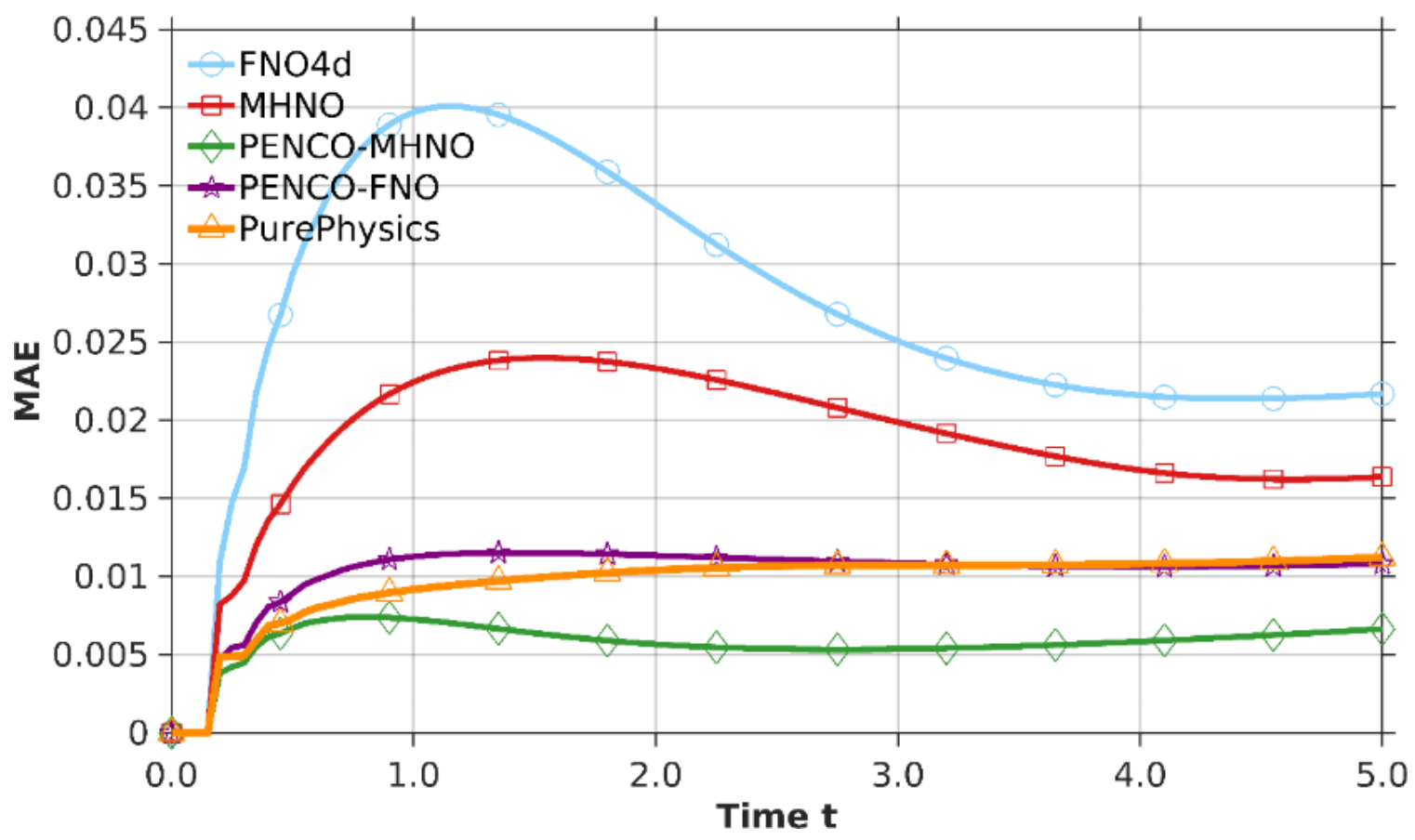

c) SH with spherical IC ($R = 5, \epsilon = 0.15, b = 2$)

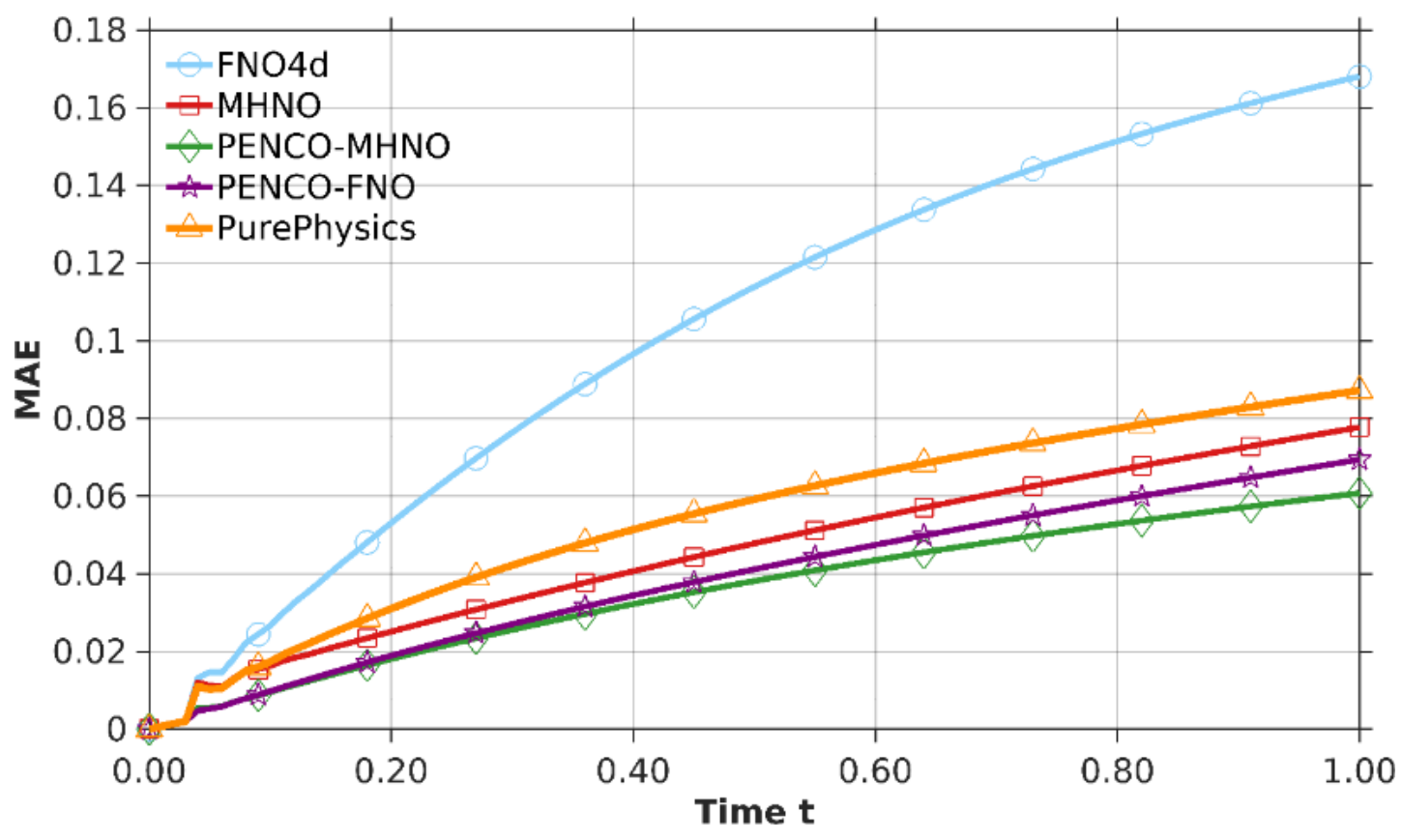

d) PFC with star-shaped IC ($\theta = \tan^{-1}(z/x), d = \sqrt{x^2 + 2y^2 + z^2}, \epsilon = 0.5$)

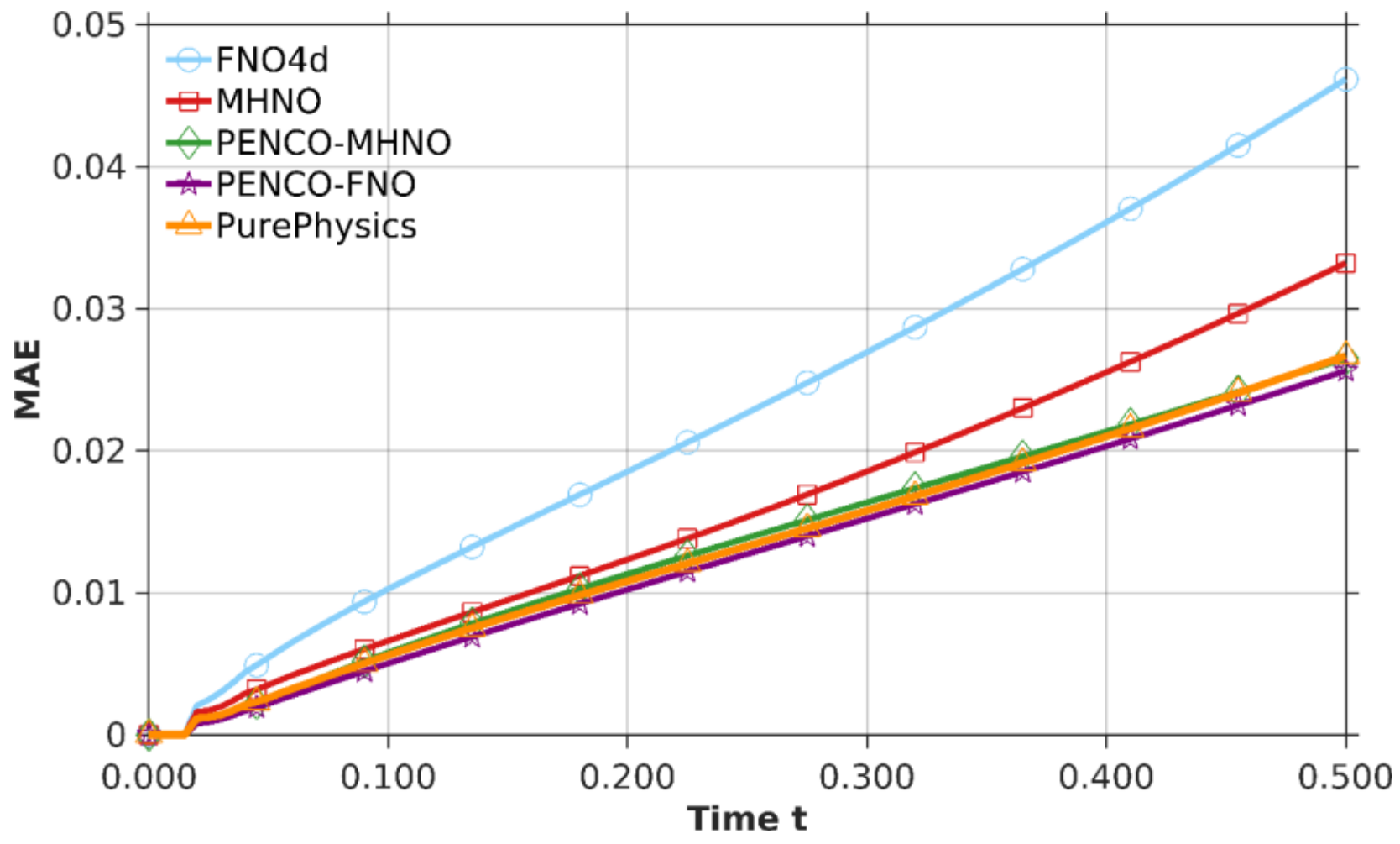

e) MBE with torus IC ($R = 0.5, r_0 = 0.1, w = 2\epsilon$)

Figure 4-16(a–e): Each panel shows MAE over time for all architectures (FNO-4D, MHNO, PENCO-MHNO, PENCO-FNO, physics-only), illustrating the effect of geometry-dependent shifts in curvature and interface structure on predictive accuracy.

## 5 Conclusion

This work introduced PENCO, a physics-guided operator learning framework for nonlinear and higher order phase-field systems. Beyond coupling the structure of the governing equations with data-driven refinement, PENCO incorporates several key innovations in its loss design such as Gauss Lobatto collocation points that enforce symmetric and stable residuals of the partial differential equation, epoch dependent weighting schedules that smoothly shift the training focus from local scheme accuracy to global spectral stability, and a low frequency anchor term that prevents drift in the dominant Fourier modes.

Across all equations considered, PENCO provides the most stable long horizon behavior and the lowest accumulated error. The hybrid formulation consistently outperforms purely data-driven models, which tend to drift once the dynamics leave the statistics of the training set, and it improves over the pure physics model by recovering fine scale features that the PDE alone cannot generate.

Furthermore, PENCO achieves these improvements while relying on a much lighter computational configuration than the original MHNO study [54], which was developed for two-dimensional settings. In our three-dimensional experiments, the spatial resolution is reduced from $64^3$ to $32^3$, the operator depth from four–six layers to two, and the training horizon is reduced by a factor of ten to fifty epochs, all while preserving accuracy, stability, and robustness.

When evaluated on spherical, star shaped, and torus based initial conditions that differ strongly from the GRF distribution used for training, PENCO maintains coherent evolution and low error for every system except CH, where the pure physics model aligns slightly more closely with the reference due to its strictly conservative structure. In the remaining equations, the hybrid model achieves the highest accuracy, demonstrating that combining physical constraints with learned spatial refinement offers strong robustness to unseen geometries.

**Future directions**

The framework can be extended in two natural directions. First, integrating fracture and crack growth models would enable prediction of crack initiation and propagation within the same hybrid operator formulation. Second, supporting non periodic boundary conditions would allow application to more realistic geometries and multi physics settings. These developments would further expand the applicability of machine learning accelerated phase-field modeling.

## Acknowledgments

The authors acknowledge funding from the German Academic Exchange Service (DAAD), which provided a doctoral scholarship to Mostafa Bamdad that supported the work presented in this paper.

# 6 Appendix: Numerical Solver

The Fourier-spectral method is a powerful numerical approach for solving phase-field equations, offering spectral accuracy by leveraging the efficiency of the Fast Fourier Transform (FFT). Below, we describe its adaptation to three-dimensional (3D) simulations, focusing on the CH equation as a representative example.

Consider a 3D periodic domain $\Omega = [l_x, r_x] \times [l_y, r_y] \times [l_z, r_z]$, discretized into a grid of $N_x \times N_y \times N_z$, and the grid spacing is defined by $h_x = \frac{L_x}{N_x}$, $h_y = \frac{L_y}{N_y}$, $h_z = \frac{L_z}{N_z}$, with domain lengths $L_x = r_x - l_x$, $L_y = r_y - l_y$ and $L_z = r_z - l_z$. The spatial coordinates are:

$$(x_m, y_n, z_l) = (l_x + mh_x, l_y + nh_y, l_z + lh_z), \qquad (0 \le m \le N_x, 0 \le n \le N_y, 0 \le l \le N_z) \tag{33}$$

The 3D Discrete Fourier Transform (DFT) of a time-varying field $\phi(x,y,z,t)$ at time step $t_k = k\Delta t$:

$$\widehat{\phi_k^{p,q,r}} = \sum_{m=1}^{N_x}\sum_{n=1}^{N_y}\sum_{l=1}^{N_z} \phi_k^{m,n,l}\ e^{-i(\xi_p x_m + \eta_q y_n + \zeta_r z_l)} \tag{34}$$

Where the wavenumbers $\xi_p = \frac{2\pi p}{L_x}$, $\eta_q = \frac{2\pi q}{L_y}$, $\zeta_r = \frac{2\pi r}{L_z}$, are indexed by integers $p, q, r$ in the ranges:

$$-\frac{N_x}{2} + 1 \le p \le \frac{N_x}{2}, \quad -\frac{N_y}{2} + 1 \le q \le \frac{N_y}{2}, \quad -\frac{N_z}{2} + 1 \le r \le \frac{N_z}{2} \tag{35}$$

The inverse discrete Fourier transform (IDFT) bridges Fourier space and physical space, synthesizing the field $\phi(x,y,z,t_k)$ from its spectral coefficients. At each time step $t_k$ the solution is reconstructed by summing contributions from all Fourier modes:

$$\phi_k^{m,n,l} = \frac{1}{N_x N_y N_z} \sum_{p=-N_x/2+1}^{N_x/2}\ \sum_{q=-N_y/2+1}^{N_y/2}\ \sum_{r=-N_z/2+1}^{N_z/2} \widehat{\phi_k^{p,q,r}}\ e^{i(\xi_p x_m + \eta_q y_n + \zeta_r z_l)} \tag{36}$$

This operation combines all wavenumber modes $(\xi_p, \eta_q,\ \zeta_r\ )$, weighted by their amplitudes $\widehat{\phi_k^{p,q,r}}$, to reproduce the field's spatial structure.

Derivatives are computed with spectral accuracy by leveraging the Fourier transform's unique properties. In Fourier space, differentiation simplifies to multiplication by wavenumbers:

$$\partial\phi/\partial x \rightarrow^{(DFT)} i\xi_p \phi^{p,q,r},\ \partial\phi/\partial y \rightarrow^{(DFT)} i\eta_q \phi^{p,q,r},\ \partial\phi/\partial z \rightarrow^{(DFT)} i\zeta_r \phi^{p,q,r} \tag{37}$$

The Laplacian acts as a low-pass filter, damping high-wavenumber modes more strongly due to the quadratic dependence on $(\xi_p, \eta_q,\ \zeta_r\ )$:

$$\Delta\phi \rightarrow^{(DFT)} -\left(\xi_p^2 + \eta_q^2 + \zeta_r^2\right)\phi^{p,q,r} \tag{38}$$

The CH equation describes phase separation dynamics and is solved using a linearly stabilized splitting scheme. The time-discretized form on a 3D grid $x_m, y_n, z_l$ in physical space is:

$$\frac{\phi_k^{m,n,l} - \phi_{k-1}^{m,n,l}}{\Delta t} = \Delta\left(2\phi_k^{m,n,l} - \epsilon^2(\Delta\phi_k)^{m,n,l} + f\left(\phi_{k-1}^{m,n,l}\right)\right), \tag{39}$$

Where $f(\phi) = \phi^3 - 3\phi$ is the nonlinear term, and $\epsilon$ controls interfacial energy. Transforming to Fourier space yields:

$$\frac{\widehat{\phi_k^{p,q,r}} - \widehat{\phi_{k-1}^{p,q,r}}}{\Delta t} = -(\xi_p^2 + \eta_q^2 + \zeta_r^2)\left(2\widehat{\phi_k^{p,q,r}} + \epsilon^2(\xi_p^2 + \eta_q^2 + \zeta_r^2)\widehat{\phi_k^{p,q,r}} + \widehat{f_{k-1}^{p,q,r}}\right), \tag{40}$$

where $\xi_p . \eta_q$ and $\zeta_r$ are wavenumbers in $x-, y-, z-$directions. Solving for $\widehat{\phi_k^{p,q,r}}$ gives the update rule:

$$\widehat{\phi_k^{p,q,r}} = \frac{\widehat{\phi_{k-1}^{p,q,r}} - \Delta t(\xi_p^2 + \eta_q^2 + \zeta_r^2)\widehat{f_{k-1}^{p,q,r}}}{1 + \Delta t\left[2(\xi_p^2 + \eta_q^2 + \zeta_r^2) + \epsilon^2(\xi_p^2 + \eta_q^2 + \zeta_r^2)^2\right]} \tag{41}$$

The physical-space solution is reconstructed via the inverse DFT:

$$\phi_k^{m,n,l} = \frac{1}{N_x N_y N_z} \sum_{p=-N_x/2+1}^{N_x/2} \sum_{q=-N_y/2+1}^{N_y/2} \sum_{r=-N_z/2+1}^{N_z/2} \widehat{\phi_k^{p,q,r}}\, e^{i(\xi_p x_m + \eta_q y_n + \zeta_r z_l)} \tag{42}$$

This method ensures high-fidelity simulations of phase-field dynamics in 3D. For further implementation details, see [7, 54].

# Supplementary Information

## S1 Supplementary Method

### S1.1 Factorized Fourier Neural Operator (FFNO) Backbone

The Factorized Fourier Neural Operator was originally introduced to reduce the complexity of Fourier based neural operators by exploiting separability in the spectral domain. Instead of learning a fully coupled multidimensional Fourier kernel, the method factorizes the operator into independent spectral components acting along each spatial dimension, which substantially reduces the number of learnable parameters while retaining long range interaction capability. This idea was formalized in [67], where the authors demonstrated that separable spectral representations combined with improved residual connections can scale efficiently to high dimensional partial differential equations while maintaining competitive accuracy.

In its standard three-dimensional form, the FFNO approximates the Fourier operator by summing independent one-dimensional spectral operators along each coordinate direction. Let $\boldsymbol{x}$ denote the input field and $K$ the learned operator. The standard FFNO representation is given by

$$K(x) = K_x(x) + K_y(x) + K_z(x), \tag{43}$$

where $K_x, K_y$, and $K_z$ denote spectral convolutions applied independently along the $x, y$, and $z$ directions. This factorization dramatically reduces model complexity and parameters compared to fully coupled FNO.

While this formulation is effective for many smooth dynamical systems, phase-field equations present additional challenges. These equations exhibit strong cross dimensional coupling and sharp interfacial structures governed by curvature driven dynamics. In such cases, the purely separable spectral approximation underrepresents interactions between spatial directions and under resolves high frequency content near interfaces. This limitation manifests as increased one step prediction error, which compounds rapidly during autoregressive rollout and leads to a sharp degradation of accuracy at later times.

To address this limitation while preserving the efficiency of the FFNO framework, we augmented the factorized operator with additional structural components designed to better reflect the physics of phase-field models. First, we introduced pairwise spectral coupling terms that explicitly model interactions between spatial directions in Fourier space. Second, we added a local three-dimensional convolutional path that provides a stencil like inductive bias, which is essential for capturing Laplacian dominated and curvature driven behavior. Third, we incorporated learnable gating mechanisms that allow the network to adaptively balance global spectral interactions and local corrections based on the evolving solution state. The resulting operator takes the following form

$$K(x) = \alpha_x K_x(x) + \alpha_y K_y(x) + \alpha_z K_z(x) + \alpha_{xy} K_{xy}(x) + \alpha_{xz} K_{xz}(x) + \alpha_{yz} K_{yz}(x) + \alpha_{\text{loc}} K_{\text{local}}(x) \tag{44}$$

where $K_{xy}, K_{xz}, K_{yz}$ denote pairwise spectral operators acting on joint frequency planes, $K_{\text{local}}$ denotes the local convolutional operator, and the coefficients $\alpha$ are learnable and normalized to ensure stable training. This formulation preserves the factorized philosophy of FFNO while significantly enhancing its expressive power for multiscale and interface dominated dynamics. The implementation of the developed FFNO backbone and all associated experiments are publicly available at https://github.com/MBamdad/PENCO/tree/Revision.

The standard FFNO backbone's result is not reported in Table S2.5 because its prediction error for phase-field equations was excessively large and not suitable for meaningful comparison. In contrast, the developed FFNO backbone demonstrates a clear improvement in predictive accuracy. In purely data-driven settings, the developed FFNO consistently outperforms the standard FNO backbone across all considered phase-field models, with especially pronounced gains observed for the MBE equation. When incorporated into the developed PENCO hybrid framework, the FFNO backbone achieves accuracy comparable to the standard FNO backbone for the AC and CH equations, while exhibiting reduced accuracy for the remaining cases. This observation suggests that, although the developed FFNO is highly effective in data-driven formulations, its factorized spectral structure may be less compatible with physics injection strategies for certain phase-field equations that require strongly coupled spatial representations.

# S2 Supplementary Tables

## S2.1 Summary of notation

Table S2.1 provides a compact reference for all mathematical symbols and model components appearing in the paper, grouped by their functional roles.

Table S2.1. Notation overview for fields, operators, and model components

| Symbol | Description |
|---|---|
| **Domains, Coordinates, Fields** | |
| $\Omega \subset R^3, t \in [0,T], \boldsymbol{x} = (x,y,z)$ | Periodic spatial domain, Time interval, Spatial coordinate |
| $u(x,t), u^n \approx u(\boldsymbol{x}, t_n)$ | Continuous field and its discrete-time representation |
| $\hat{u}^{n+1}, \Delta t$ | Predicted next-step field, and uniform time step |
| **Differential Operators** | |
| $\nabla u, \nabla^2 u, \nabla^4 u$ | Gradient of $u$, Laplacian, Biharmonic operator |
| $\mathcal{R}, \mathcal{D}, \mathcal{M}, L(u), \mathcal{N}(u)$ | PDE residual, evolution operator, mobility operator, linear and nonlinear part of the PDE |
| **Neural Operators** | |
| $\mathcal{G}_\theta, a(\boldsymbol{x})$ | Learned solution operator and its input initial condition |
| $v_\tau, \mathcal{W}_\tau, K_\tau$ | Latent feature, local linear transform, spectral convolution at layer $\tau$ |
| $\mathcal{P}, \mathcal{Q}, R_\tau(\boldsymbol{k}), \boldsymbol{k} = (k_x, k_y, k_z)$ | Lifting & projection networks, spectral filter, wavenumber vector |
| **MHNO-specific symbols** | |
| $\mathcal{Q}_n,\ \mathcal{H}_n,\ v_a(x)$ | Projection head, transition head, and shared spatial latent feature |
| **PENCO** | |
| $\tau_{1,2} = \frac{1}{2} \pm \frac{1}{2\sqrt{5}}, R_\tau, \widetilde{\mathrm{R}}_\tau$ | Gauss–Lobatto nodes, collocation residual, and normalized residual |
| $\dot{u^n} = \frac{\hat{u}^{n+1} - u^n}{\Delta t}, u_\tau = (1-\tau)u^n + \tau\hat{u}^{n+1}$ | Predicted time derivative and interpolated midpoint state |
| $E[u], F(u), \epsilon$ | Free-energy functional, bulk potential, interfacial thickness |
| $u_{SI}^{n+1}, \mathcal{P}_{low-k}(u), \lambda, \alpha$ | Semi-implicit update, low-frequency spectral projection, and physics–data weighting/loss balance factors |
| PENCO-FNO, PENCO-MHNO | FNO and MHNO architectures augmented with PENCO physics constraints |

## S2.2 Summary of hyperparameters

Table S2.2 summarizes the hyperparameters used across all neural operator models considered in this study, including FNO, MHNO, PENCO-FNO, PENCO-MHNO, and the pure-physics model. The table reports the number of training epochs, dataset sizes, learning rates, number of layers, model widths and depths, and the number of retained Fourier modes.

The temporal hyperparameters $T_{in}$ and $T_{out}$ control how the models process time. The first sets how many past solution states are provided as input, and the second determines how many future states are predicted in each forward pass.

Table S2.2: Summary of neural operator hyperparameters used for all experiments

| Problems | Epochs | $N_{train}$ | $N_{test}$ | $T_{in}, T_{out}$ | Lr | $W_Q$ | $W_H$ | M | W | $N_l$ | $N_Q$ | $N_H$ |
|---|---|---|---|---|---|---|---|---|---|---|---|---|
| **AC** | 50 | 50,100,200 | 50 | 4, 1 | 0.001 | 10 | 10 | 10 | 10 | 2 | 2 | 2 |
| **CH** | 50 | - | - | - | 0.001 | 10 | 10 | 10 | 10 | 2 | 2 | 2 |
| **SH** | 150 | - | - | - | 0.001 | 10 | 10 | 10 | 10 | 2 | 2 | 2 |
| **PFC** | 50 | - | - | - | 0.001 | 10 | 10 | 10 | 10 | 2 | 2 | 2 |
| **MBE** | 50 | - | - | - | 0.0005 | 11 | 11 | 10 | 10 | 2 | 2 | 2 |

## S2.3 Summary of Computational Performance Metrics

Supplementary table S2.3 details the computational performance of our models, reporting for each governing equation the discretized domain resolution, the total number of dataset samples, the total time required for dataset generation, the training duration for datasets of 50, 100, and 200 samples, and the inference time for a single temporal rollout ($N_t = 100$, using the PENCO–MHNO model with $\lambda = 0.25$).

Table S2.3. Computational performance metrics for the evaluated equations

| **Examples** | Resolution | Samples | Dataset (s) | Training (s) | | | Inference (s) | | |
|---|---|---|---|---|---|---|---|---|---|
| | | | | **50** | **100** | **200** | **50** | **100** | **200** |
| **AC** | 32×32×32 | 250 | 272.5 | 13.6 | 35.2 | 61.7 | 0.22 | 0.29 | 0.19 |
| **CH** | " | " | 246.5 | 104.7 | 164.6 | 258.1 | 0.37 | 0.26 | 0.22 |
| **SH** | " | " | 270 | 64.0 | 112.4 | 196.6 | 0.27 | 0.42 | 0.41 |
| **PFC** | " | " | 257.8 | 39.7 | 83.5 | 159.0 | 0.30 | 0.36 | 0.31 |
| **MBE** | " | " | 315.4 | 185.7 | 351.8 | 641.6 | 0.4 | 0.35 | 0.31 |

## S2.4 Comparative Overview of Operator-Learning Frameworks

This section presents a high-level comparison of the operator-learning frameworks considered in this study, focusing on their treatment of physics constraints, temporal structure, numerical scheme consistency, and data dependence, as summarized in Table S2.4. The comparison highlights the conceptual differences between purely data-driven neural operators, existing physics-informed approaches, and the proposed PENCO framework.

Table S2.4. High-level comparison of operator-learning frameworks.

| Method | Operator Type | Physics Enforcement | Temporal Treatment | Numerical Scheme Consistency | Data Dependence |
|---|---|---|---|---|---|
| **FNO-4D** | Neural operator | None | Fully data-driven | No | High |
| **PINO** | " | PDE residual penalties | Continuous or weakly discretized | No | Medium |
| **MHNO** | " | None | Discrete, step-aware | No | High |
| **VINO** | " | Variational energy residuals | Continuous | No | Medium |
| **PENCO** | " | Collocation, energy, spectral stability | IMEX-aligned temporal collocation | Yes | Low |

## S2.5 Numerical Results for FFNO Backbone

When FFNO-4D is adopted as the backbone architecture, the network is configured with 12 Fourier modes, a feature width of 12, four FFNO layers, and an FFNO feedforward expansion factor of 2 (set to 6 for the CH case). All other hyperparameters are taken to be identical to those listed in Table S2.2. Table S2.5 reports the corresponding prediction errors at selected rollout times and the total training cost for FFNO-based models across the considered phase-field equations.

Table S2.5. Prediction errors and training time for FFNO-based architectures.

| | N | Method | $0\Delta t$ | $50\Delta t$ | $100\Delta t$ | Total training time (min.) |
|---|---|---|---|---|---|---|
| AC | 50 | FFNO-4d | 0 | 0.16 | 0.23 | 1.93 |
| | 50 | MHNO-FFNO | 0 | 0.19 | 0.24 | 0.99 |
| | 50 | PENCO-FFNO ($\lambda = 0.25$) | 0 | 0.06 | 0.07 | 2.22 |
| | 50 | PENCO-MHNO_FFNO ($\lambda = 0.25$) | 0 | 0.08 | 0.1 | 1.55 |
| | 200 | FFNO-4d | 0 | 0.05 | 0.06 | 5.87 |
| | 200 | MHNO-FFNO | 0 | 0.08 | 0.09 | 4.22 |
| | 200 | PENCO-FFNO ($\lambda = 0.25$) | 0 | 0.02 | 0.02 | 6.04 |
| | 200 | PENCO-MHNO_FFNO ($\lambda = 0.25$) | 0 | 0.02 | 0.03 | 4.09 |
| CH | 50 | FFNO-4d | 0 | 0.14 | 0.20 | 6.65 |
| | 50 | MHNO-FFNO | 0 | 0.14 | 0.21 | 7.13 |
| | 50 | PENCO-FFNO ($\lambda = 0.25$) | 0 | 0.08 | 0.13 | 9.27 |
| | 50 | PENCO-MHNO_FFNO ($\lambda = 0.25$) | 0 | 0.08 | 0.13 | 6.73 |
| | 200 | FFNO-4d | 0 | 0.09 | 0.14 | 31.66 |
| | 200 | MHNO-FFNO | 0 | 0.12 | 0.17 | 32.61 |
| | 200 | PENCO-FFNO ($\lambda = 0.25$) | 0 | 0.03 | 0.04 | 35.7 |
| | 200 | PENCO-MHNO_FFNO ($\lambda = 0.25$) | 0 | 0.05 | 0.07 | 24.96 |
| SH | 50 | FFNO-4d | 0 | 0.14 | 0.21 | 2.4 |
| | 50 | MHNO-FFNO | 0 | 0.16 | 0.25 | 2.17 |
| | 50 | PENCO-FFNO ($\lambda = 0.25$) | 0 | 0.07 | 0.12 | 3.28 |
| | 50 | PENCO-MHNO_FFNO ($\lambda = 0.25$) | 0 | 0.1 | 0.16 | 2.39 |
| | 200 | FFNO-4d | 0 | 0.1 | 0.16 | 13.9 |
| | 200 | MHNO-FFNO | 0 | 0.13 | 0.19 | 10.46 |
| | 200 | PENCO-FFNO ($\lambda = 0.25$) | 0 | 0.04 | 0.09 | 13.82 |
| | 200 | PENCO-MHNO_FFNO ($\lambda = 0.25$) | 0 | 0.06 | 0.11 | 10.01 |
| PFC | 50 | FFNO-4d | 0 | 0.15 | 0.26 | 2.7 |
| | 50 | MHNO-FFNO | 0 | 0.18 | 0.32 | 1.99 |
| | 50 | PENCO-FFNO ($\lambda = 0.25$) | 0 | 0.07 | 0.12 | 2.48 |
| | 50 | PENCO-MHNO_FFNO ($\lambda = 0.25$) | 0 | 0.19 | 0.21 | 1.8 |
| | 200 | FFNO-4d | 0 | 0.1 | 0.17 | 9.06 |
| | 200 | MHNO-FFNO | 0 | 0.09 | 0.15 | 6.12 |
| | 200 | PENCO-FFNO ($\lambda = 0.25$) | 0 | 0.03 | 0.06 | 7.85 |
| | 200 | PENCO-MHNO_FFNO ($\lambda = 0.25$) | 0 | 0.06 | 0.11 | 6.44 |
| MBE | 50 | FFNO-4d | 0 | 0.31 | 0.24 | 6.38 |
| | 50 | MHNO-FFNO | 0 | 0.5 | 0.53 | 5.58 |
| | 50 | PENCO-FFNO ($\lambda = 0.25$) | 0 | 0.08 | 0.13 | 4.43 |
| | 50 | PENCO-MHNO_FFNO ($\lambda = 0.25$) | 0 | 0.1 | 0.19 | 5.55 |
| | 200 | FFNO-4d | 0 | 0.41 | 0.92 | 19.46 |
| | 200 | MHNO-FFNO | 0 | 0.34 | 0.31 | 21.58 |
| | 200 | PENCO-FFNO ($\lambda = 0.25$) | 0 | 0.06 | 0.15 | 25.34 |
| | 200 | PENCO-MHNO_FFNO ($\lambda = 0.25$) | 0 | 0.06 | 0.15 | 22.13 |

# S3 Supplementary Plots

## S3.1 Performance and Robustness of FNO-Based Models

This three-dimensional box plot (Figure S3-1) summarizes the distribution of final time relative $L_2$ errors across all equations, with the horizontal axes indexing the different operator architectures and training set sizes and the vertical axis showing the error magnitude. Each box aggregates results over the benchmark problems, so the median, interquartile range, and whiskers provide a compact view of how accuracy and variability change with method choice and data availability.

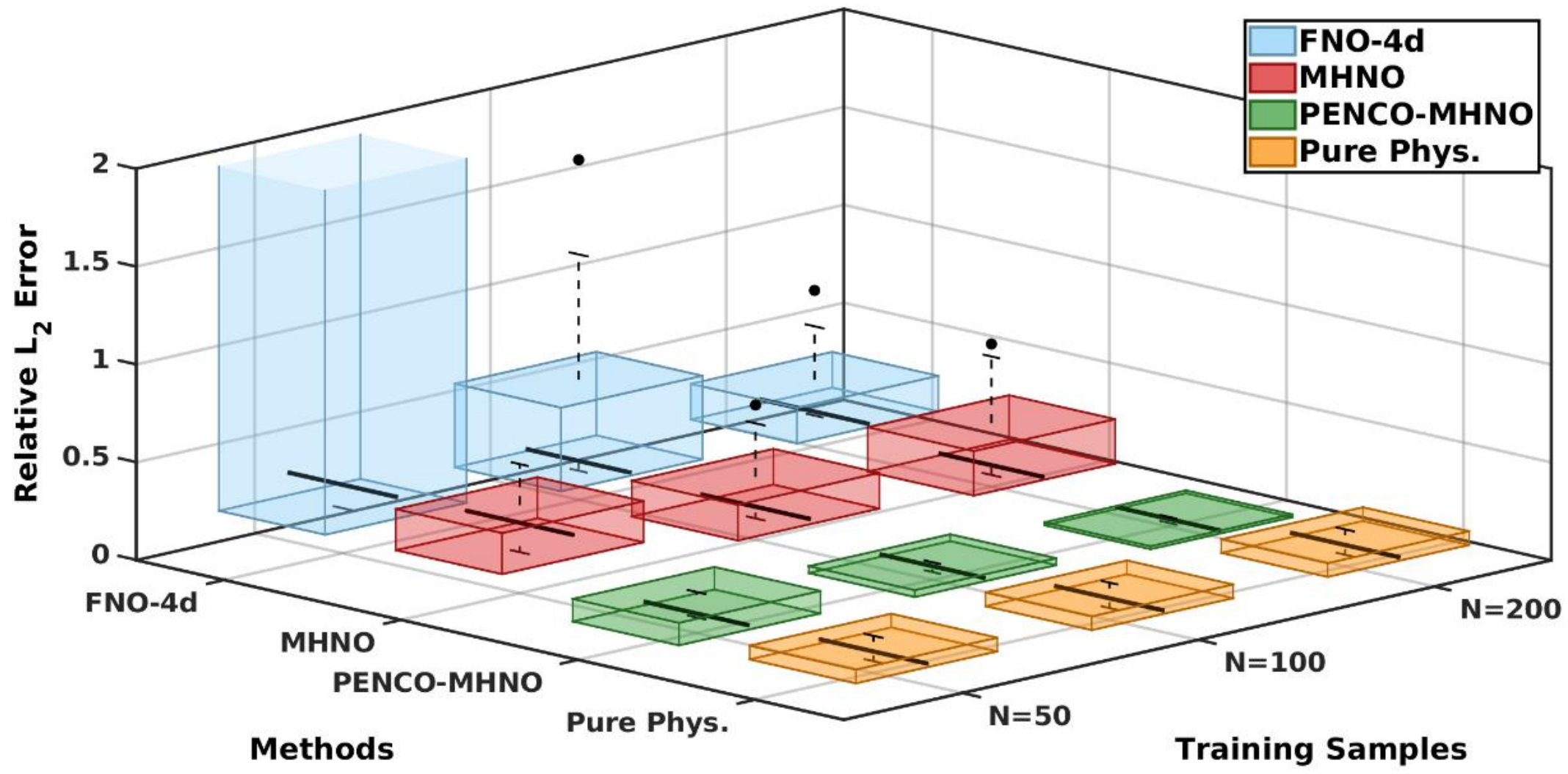

Figure S3-1: 3D box plot illustrating accuracy and variability of each operator at the final rollout time for multiple training-set sizes.

## S3.2 Performance and Robustness of FFNO-Based Models

The box plot Figure S3-2 summarize the variability of the relative $L_2$ error across the five phase-field equations at $100\Delta t$. While increasing the training samples from $N = 50$ and $N = 200$ generally reduces the median error, the box height reveals how consistently each method performs across different equations. In particular, FFNO exhibits a larger box at $N = 200$, indicating increased variability due to uneven improvements across equations, despite better accuracy for some cases. In contrast, PENCO-FFNO and PRNCO-MHNO_FFNO show both lower medians and smaller boxes, demonstrating not only improved accuracy but also more robust and consistent performance across all equations.

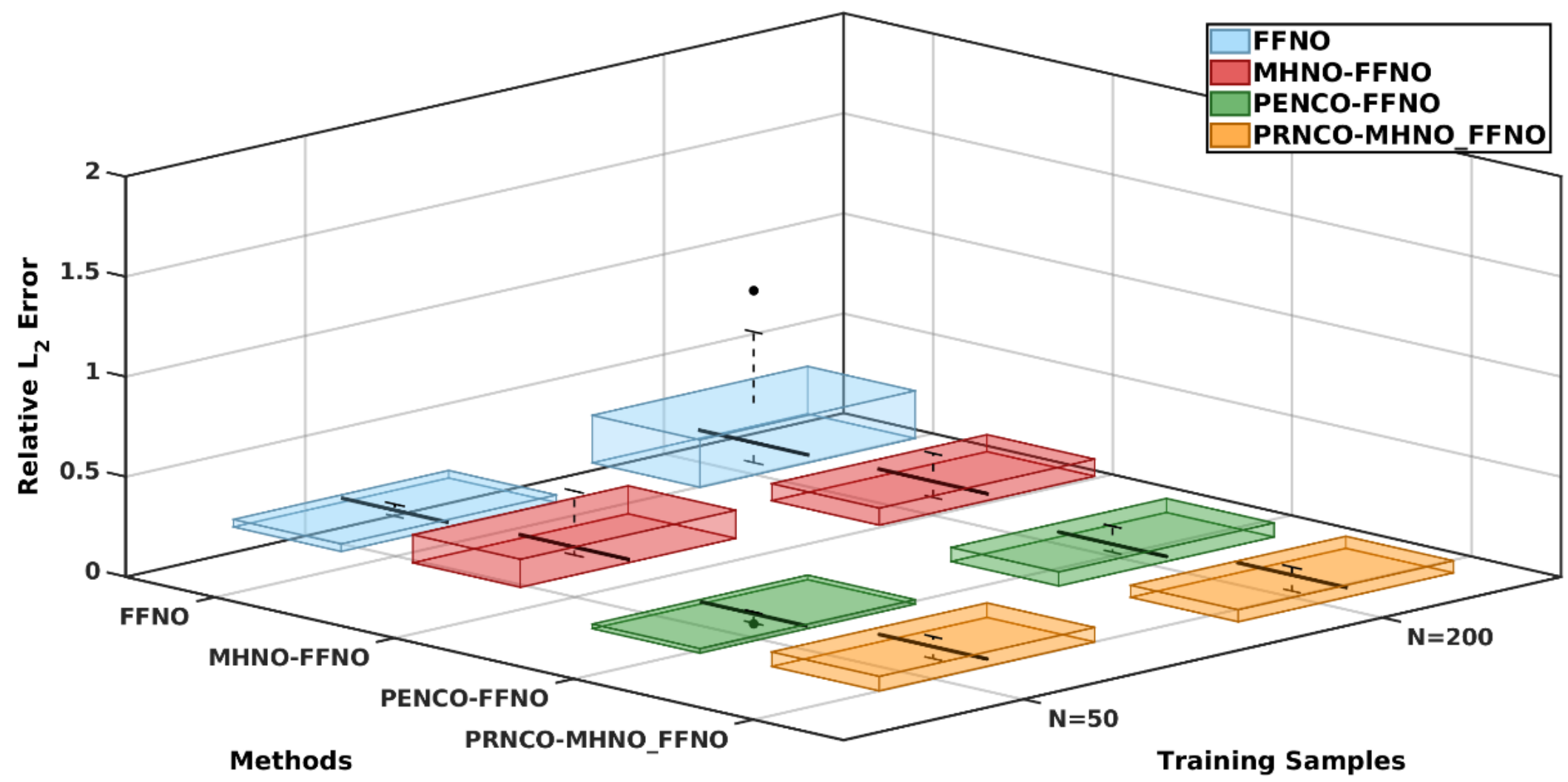

Figure S3-2: Comparison of accuracy and cross-equation robustness of different methods

### S3.3 Physical Consistency Diagnostics

In addition to the normalized L2 error, we now report physically motivated diagnostics to assess long-term stability and consistency. Specifically, we include energy evolution and mass conservation error for representative cases, which confirm the expected physical behavior over long-time horizons (Figure S6-3).

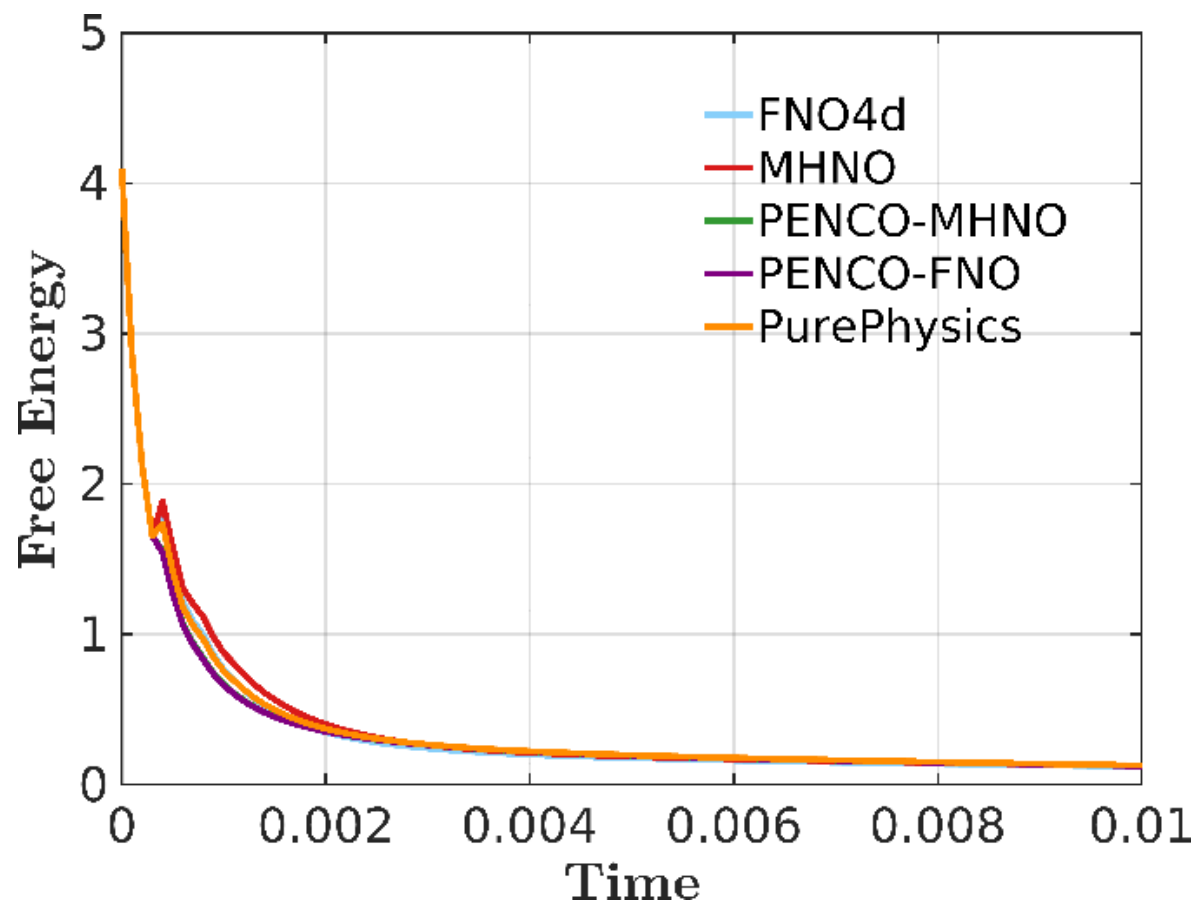

a) AC energy vs. time

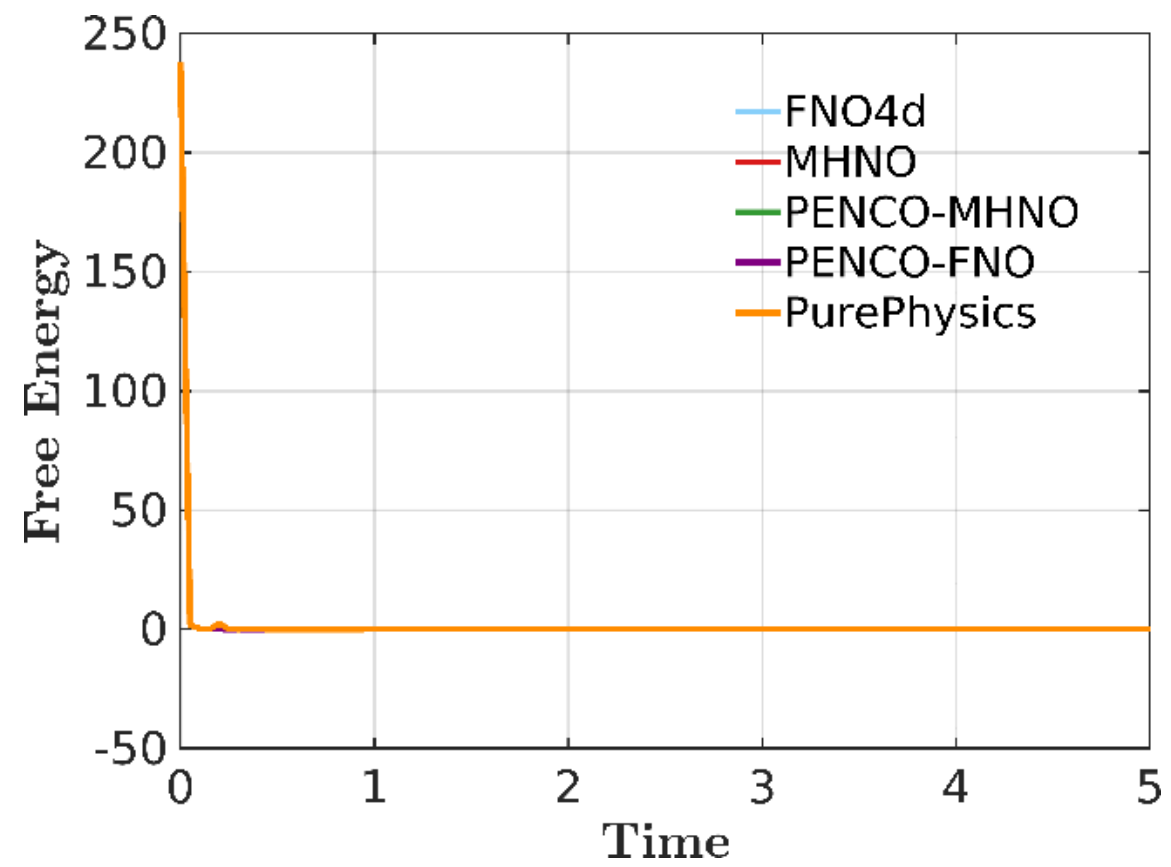

b) SH energy vs. time

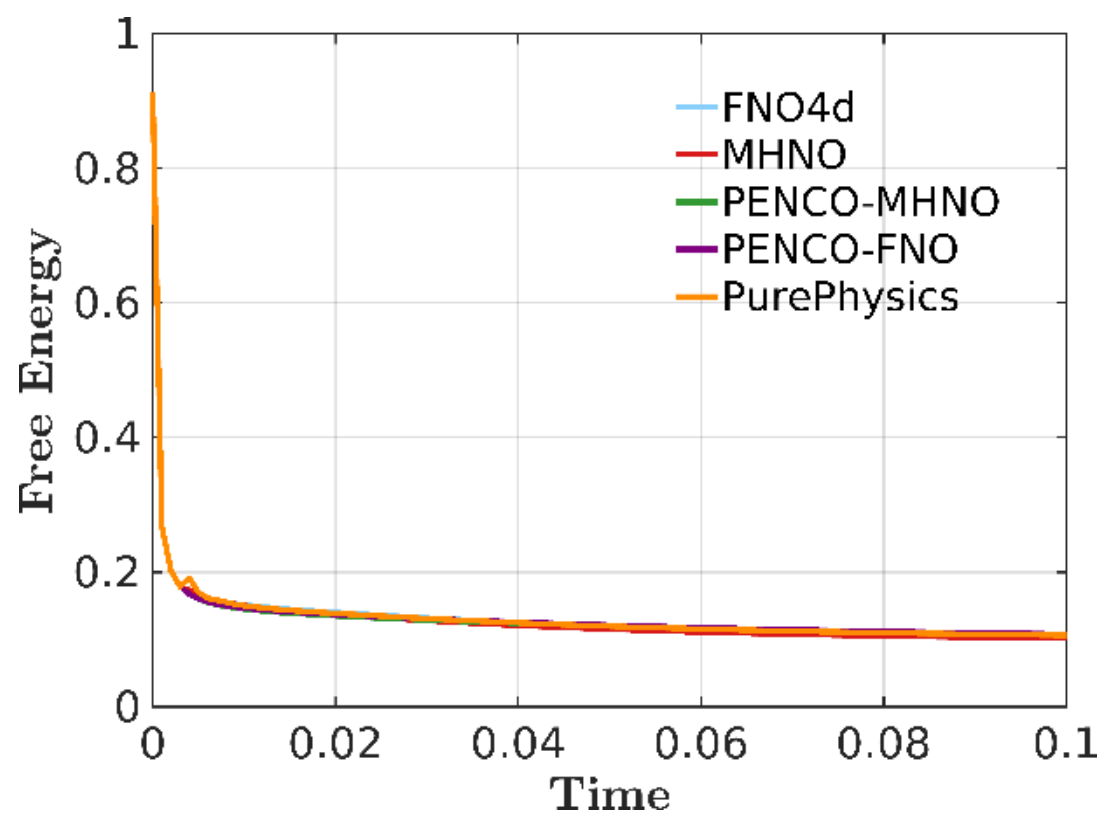

c) CH energy vs. time

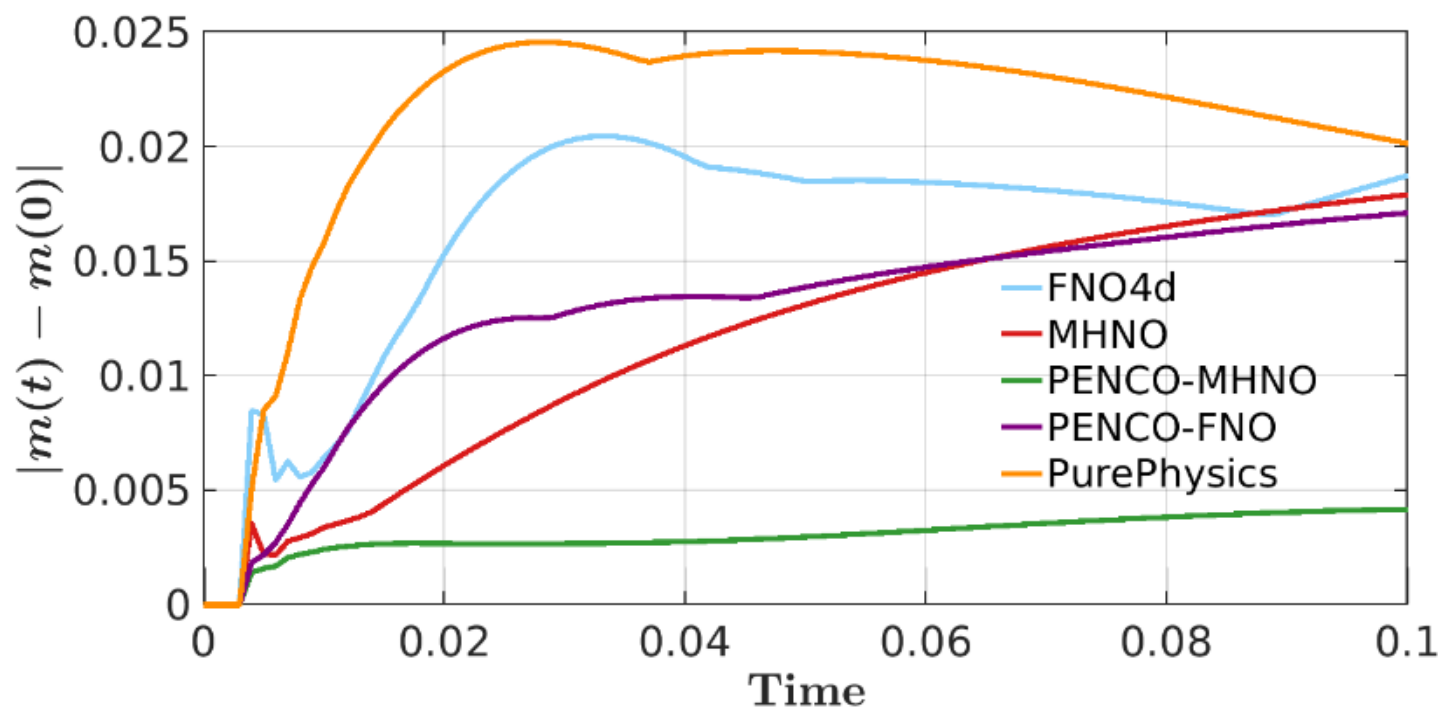

d) CH mass conservation

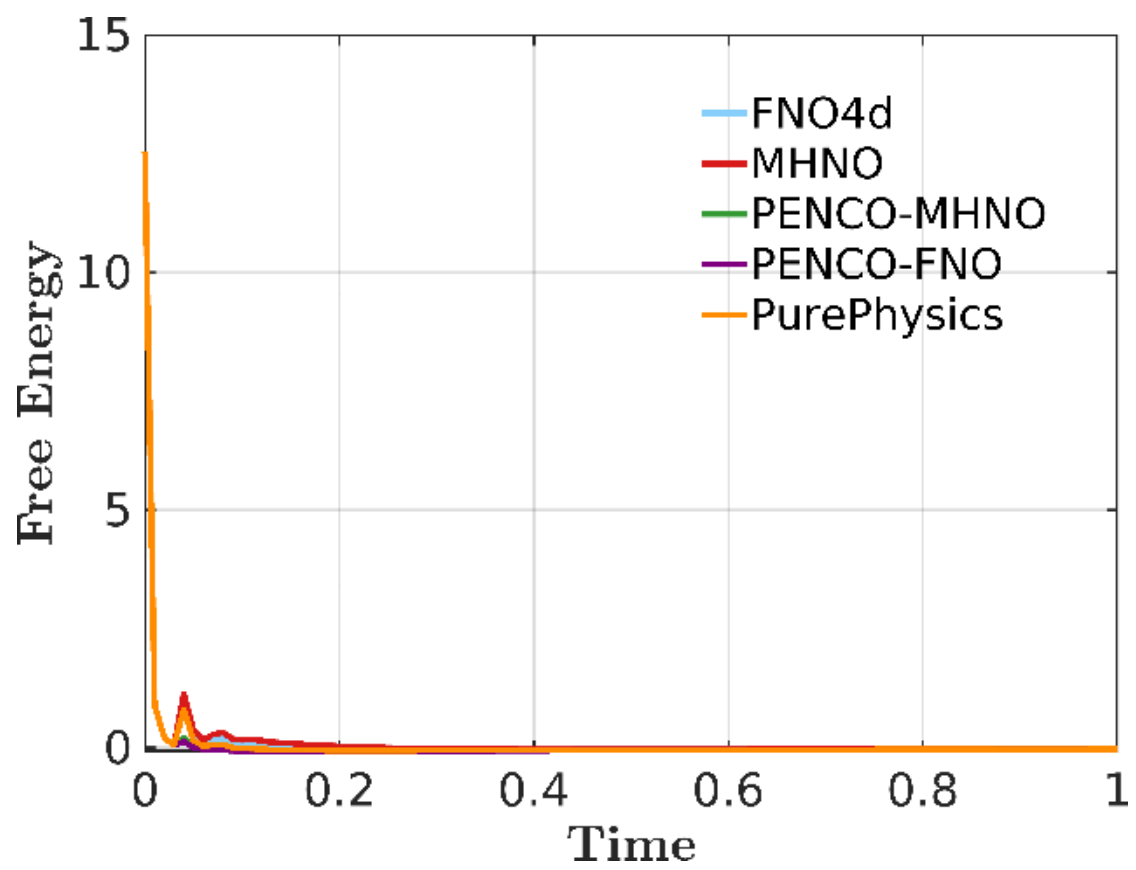

e) PFC energy vs. time

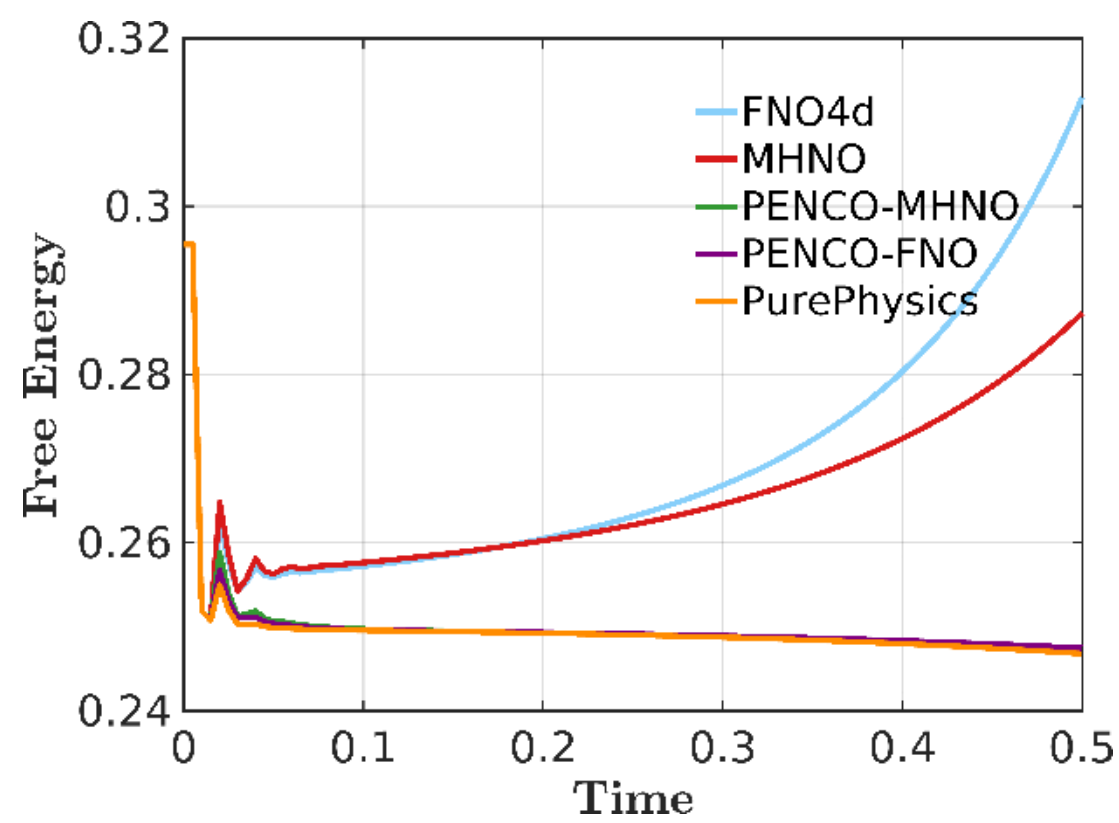

f) MBE energy vs. time

Figure S6-3: Energy evolution and mass conservation error (only for CH case) for representative test cases, demonstrating long-term physical stability.